\documentclass[pra,aps,twocolumn]{revtex4}
\usepackage{times,epsfig,amssymb,amsfonts,amsmath,amsthm}%\usepackage{epsfig}
\newcommand{\ket}[1]{|#1\rangle}
\newcommand{\bra}[1]{\langle #1|}
\newcommand{\proj}[1]{\ket{#1}\bra{#1}}
\newcommand{\tr}{\text{Tr}}
\usepackage{graphicx}
\usepackage{subfigure}
\usepackage{xcolor} %DIF >
\usepackage{upgreek}
\usepackage[figuresright]{rotating}
\usepackage{hyperref}

\hypersetup{
colorlinks=true,
linkcolor=blue,
urlcolor=blue,
citecolor=blue}

\begin{document}

\title{Sequential quantum nonlocality sharing under local noisy quantum channels}

\author{Na Li$^{1,2}$}
\author{Chen-Yue Li$^{1}$}
\author{Yu-Hong Zheng$^3$}
\email{zhengyh@sjzpc.edu.cn}
\author{Wen-Long Ma$^{2,4}$}
\email{wenlongma@semi.ac.cn}
\author{Li-Hang Ren$^{1}$}
\email{renlihang@hebtu.edu.cn}
\author{Yan-Kui Bai$^{1}$}
\email{ykbai@semi.ac.cn}

\affiliation{$^1$College of Physics and Hebei Key Laboratory of Photophysics Research and Application, Hebei Normal University, Shijiazhuang, Hebei 050024, China\\
$^2$ State Key Laboratory of Semiconductor Physics and Chip Technologies, Institute of Semiconductors, Chinese Academy of Sciences, Beijing 100083, China\\
$^3$ Shijiazhuang Posts and Telecommunications Technical College, Shijiazhuang, Hebei 050021, China\\
$^4$ Center of Materials Science and Opto-Electronic Technology, University of Chinese Academy of Sciences, Beijing 100049, China}

\begin{abstract} 
Sequential sharing of quantum nonlocality (SSQN) is crucial for device-independent tasks in quantum information processing, wherein relaying the post-measurement qubit through a local quantum channel to a subsequent observer constitutes an essential operational step. Here we present a theoretical analysis of noise robustness of sequential sharing for bipartite Bell and tripartite Mermin nonlocality under the influence of local phase-flip, bit-flip, and depolarizing quantum channels. It is proved that arbitrarily many independent observers can sequentially share the quantum nonlocality of Bell, Greenberger-Horne-Zeilinger, and W states via respective noise-immune channels, whereas such unbound feature of SSQN is lost under other local noisy quantum channels. Furthermore, we demonstrate that the noise-immune channel enabling unbounded SSQN can be switched by employing our newly designed measurement strategies assisted by local unitary operations on the initial entangled states. Moreover, as illustrative examples of noise robustness, we propose two concrete schemes for sharing Bell and Mermin nonlocality with two sequential local observers on one side subject to local noisy channels. Our work establishes a practical framework for realizing the SSQN under noisy quantum channels, and reveals the connection between noise robustness and measurement strategies.
\end{abstract}

\maketitle

\section{Introduction}

Quantum nonlocality is one of the most striking characteristics of quantum theory \cite{epr35pr,bohr35pr}, which yields the predictions that cannot be accounted for by any local theory \cite{jsb64,clauser72prl,aspect82prl,zeilinger98prl,panjw17sci,nds14rmp}. Along with the rapid development of quantum science and technology, quantum nonlocality has emerged as a critical resource that underpins the advantages in many quantum information processing tasks, such as device-independent quantum key distribution \cite{jb05prl,aa06prl,aa07prl,hmb12prl,bancal22nat,panjw22prl,panjw26sci}, communication complexity reduction \cite{brukner04prl,brunner09prl,buhrman10rmp}, certification of randomness \cite{sp10nat,rc12nat,lw22prl}, and so on. Conventionally, once the nonlocality test  is performed, the post-measurement state can no longer exhibit quantum nonlocality. However, a new class of protocols termed sequential sharing of quantum nonlocality (SSQN) \cite{rs15prl,pjb20prl,ren19pra,dd19pra,tgz21pra,cheng21pra,cheng22pra,ren22pra,yx23pra,bz24pra,Sun24qinp,as22prl,ssa24prl,tgz24pra,saha19qinp,wen25qinp,jg26pla} allows multiple independent observers to sequentially obtain nonlocal correlations, which has been experimentally demonstrated in various optical systems utilizing weak measurements \cite{ms17qst,mjh18njp,tff20pra,gf20pra} or projective measurements assisted by classical randomness \cite{yys24njp,xue26prl} (see also the review paper \cite{ren25physrep} and references therein).  

In the prototypical scenario of SSQN \cite{rs15prl,pjb20prl}, a single Alice aims to establish nonlocal correlations with a sequence of Bobs through a deliberate weak measurement strategy, where Alice initially shares a pair of entangled qubits with Bob$_1$ who performs a measurement and sends the post-measurement qubit to Bob$_2$ and each subsequent Bob repeats the measurement-and-sending operations. Brown and Colbeck \cite{pjb20prl} showed that arbitrarily many independent observers can sequentially share bipartite Bell nonlocality of a maximally entangled two-qubit state indicated by the Clauser-Horne-Shimony-Holt (CHSH) inequality \cite{jfc69prl}. The unbounded sharing of Bell nonlocality can also be achieved via qubit projective measurements assisted by classical randomness \cite{as22prl,ssa24prl}. Furthermore, the SSQN protocols are generalized to bipartite high-dimensional and multipartite entangled states \cite{ren25physrep}. For example, in tripartite settings, it is proved that the unbounded sequential sharing of Mermin nonlocality \cite{ndm90prl} is available for three-qubit Greenberger-Horne-Zeilinger (GHZ) and W states \cite{yx23pra,bz24pra}. In all existing  SSQN protocols, sending the post-measurement qubit from the initial independent observer to the next one in the sequence constitutes a necessary step, where it is strictly assumed that the qubit is transmitted via a noiseless quantum channel. However, realistic transmissions of the qubit are inevitably subject to channel noise. Therefore, it is of both theoretical and practical interests to study whether arbitrarily many observers are still able to sequentially share quantum nonlocality in the presence of local channel noise. Moreover, given the characteristics of noise in quantum channels, it remains unknown whether the influences of noise can be actively suppressed by a newly designed measurement strategy. 

In this paper, we study the robustness of sequential sharing of bipartite Bell-CHSH nonlocality \cite{jfc69prl} and tripartite Mermin nonlocality \cite{ndm90prl} against noise in local phase-flip, bit-flip and depolarizing quantum channels that link a sequence of independent observers. It is proved that Bell-CHSH nonlocality of an initial two-qubit Bell state can be sequentially shared by a single Alice and arbitrarily many independent Bobs in the presence of local noisy phase-flip channels via the weak measurement strategy originally presented by Brown and Colbeck \cite{pjb20prl} for the noiseless scenario, while the unbounded SSQN cannot be sustained under the influence of the other two local noisy channels. Remarkably, the noise-immune channel for sequential Bell nonlocality sharing can be switched from phase-flip channel to bit-flip channel by adopting our proposed measurement strategy. Furthermore, for tripartite settings, it is proved that the unbounded sequential Mermin nonlocality sharing of initial three-qubit GHZ and W states using the weak measurement strategies of Refs. \cite{yx23pra,bz24pra}, designed for noiseless cases, can only succeed under their specific noise-immune transmission channels. Additionally, we also propose the corresponding new measurement strategies assisted by local unitary operations, which enable switching of noise-immune local quantum channels for unbounded SSQNs. As illustrative examples of noise robustness, we present two schemes for double violations of SSQNs based on Bell and GHZ states under channel noise, employing different measurement strategies.

This paper is organized as follows. In Sec. II, we briefly recapitulate quantum nonlocality characterized by CHSH and Mermin inequalities, and introduce the typical noisy quantum channels. In Sec. III, the unbounded SSQN of a two-qubit Bell state is studied in the presence of local noisy channels. Furthermore, the robustness of SSQN with GHZ and W states against channel noise is studied in Sec. IV. Moreover, in Sec. V, to analyze noise resilience, we present two concrete schemes for the double violations of SSQN using Bell and GHZ states in the noisy scenario. Finally, we provide a discussion and conclude this paper in Sec. VI.

\section{preliminaries}

Quantum nonlocality is a fundamental feature that distinguishes quantum theory from the classical local hidden variable (LHV) model \cite{jsb64}. In a two-qubit system, two spatially separated observers, Alice and Bob, choose bipartite measurement settings $X_i$ and $Y_j$ with $i,j\in\{0,1\}$ and obtain the binary outputs $a,b\in\{0,1\}$. It is said that there is a LHV model capable of explaining the joint probability if it can be written as the following decomposition \cite{jfc69prl}
\begin{equation}\label{eq:local_model_bipartite}
\mathcal{P}(ab|X_iY_j)=\sum_{\lambda}q_{\lambda}\mathcal{P_{\lambda}}(a|X_i)\mathcal{P_{\lambda}}(b|Y_j),
\end{equation}
where $\{q_\lambda\}$ is a distribution over the hidden variable $\lambda$ with $0\le q_\lambda\le1$ and $\sum_\lambda q_\lambda=1$, and the related local probabilities are $\mathcal{P_{\lambda}}(a|X_i)=\mathcal{P}(a|X_i,\lambda)$ and $\mathcal{P_{\lambda}}(b|Y_i)=\mathcal{P}(b|Y_i,\lambda)$, respectively. The correlation function can be given from the joint probabilities $\langle X_iY_j\rangle=\sum_{a,b}(-1)^{a+b}\mathcal{P}(ab|X_iY_j)$. Then, the CHSH inequality derived from the LHV model has the form \cite{jfc69prl}  
\begin{equation}\label{eq:CHSH}
I_{\mathcal{CHSH}}=\langle X_0Y_0\rangle+\langle X_1Y_0\rangle+\langle X_0Y_1\rangle-\langle X_1Y_1\rangle\le2.
\end{equation}
This inequality can be violated up to $2\sqrt{2}$ by a two-qubit maximally entangled state under some specific measurements, which shows that the prediction of quantum mechanics is incompatible with the LHV theory. Moreover, for a generic two-qubit entangled state, the maximal expectation value of the CHSH operator can be calculated via the Horodecki parameter \cite{hor95pla}. 

Next, we consider a tripartite scenario in which three spatially separated observers, Alice, Bob, and Charlie, perform the measurements $X_i$, $Y_j$, and $Z_k$ with $i,j,k\in\{0,1\}$ on their subsystems and obtain the binary outcomes $a, b, c\in\{0,1\}$. When the joint probability correlations are called fully local, we have the relation \cite{ndm90prl}
\begin{equation}\label{eq:local_model_tripartite}
\mathcal{P}(abc|X_iY_jZ_k)=\sum_{\lambda}q_{\lambda}\mathcal{P_{\lambda}}(a|X_i)\mathcal{P_{\lambda}}(b|Y_j)\mathcal{P_{\lambda}}(c|Z_k),
\end{equation}
where the distribution over the hidden variable satisfies $0\leq q_\lambda\leq1$ and $\sum_{\lambda}q_{\lambda}=1$. The corresponding correlation function can be obtained in terms of these joint probabilities $\langle X_iY_jZ_k\rangle=\sum_{a,b,c}(-1)^{a+b+c}\mathcal{P}(abc|X_iY_jZ_k)$. Then the tripartite quantum nonlocality can be detected by the violation of the Mermin inequality \cite{ndm90prl}, which has the form
\begin{equation}\label{eq:Mermin}
\begin{aligned}
I_{\mathcal{M}}=&\langle X_1Y_0Z_0\rangle\!+\!\langle X_0Y_1Z_0\rangle\!+\!\langle X_0Y_0Z_1\rangle\!-\!\langle X_1Y_1Z_1\rangle \\
\le& 2.
\end{aligned}
\end{equation} 
In quantum theory, the maximal expectation for the Mermin operator is $4$, and any violation of the inequality exceeding $2\sqrt{2}$ certifies genuine tripartite entanglement \cite{dc02prl}. Beyond the standard Mermin nonlocality, there are other forms of tripartite nonlocality witnessed by the violations of the Svetlichny and nonsignal inequalities \cite{gs87prd,pr14Quantum,jdb13pra}.

In the SSQN protocol, transmission of the post-measurement qubit through a sequence of independent observers is an indispensable component, which is inevitably affected by noise in local quantum channels. Here, we consider typical noisy channels, such as the phase-flip, bit-flip, and depolarizing channels. Within the Kraus operator-sum formalism \cite{nielsen10cup}, a noisy quantum channel acting on an input state $\rho$ can be described by
\begin{equation}\label{channels}
\varepsilon_{i}(\rho)=\sum\limits_{m}E_m^{(i)}\rho E_m^{(i)\dagger},
\end{equation}
where $\varepsilon_{i}$ with $i=1,2,3$ denotes different channels and the operator elements $E_m$ satisfy the completeness property $\sum_m E_m^\dagger E_m=\mathbb{I}$.  Then, the phase-flip channel $\varepsilon_1$ can be expressed via two Kraus operators
\begin{eqnarray}\label{phase}
E_0^{(1)}=\sqrt{p}\mathbb{I},\quad E_1^{(1)}=\sqrt{1-p}\sigma_z,
\end{eqnarray}
where $\sigma_z$ is the Pauli $Z$ operator, and the real parameter $p$ denotes the probability that the qubit remains unchanged. The Kraus operators for the bit-flip channel $\varepsilon_2$ have the form 
\begin{eqnarray}\label{bit}
E_0^{(2)}=\sqrt{p}\mathbb{I},\quad E_1^{(2)}=\sqrt{1-p}\sigma_x,
\end{eqnarray}
where $\sigma_x$ is the Pauli $X$ operator and $p$ is the probability that the qubit remains unflipped. Moreover, the depolarizing channel $\varepsilon_3$ can be formulated by Eq. \eqref{channels} with the following four Kraus operators
\begin{eqnarray}\label{dep}
\begin{aligned}	
E_0^{(3)}=\sqrt{\frac{1+3p}{4}}\mathbb{I},\quad E_1^{(3)}=\frac{\sqrt{1-p}}{2}\sigma_x, \\
E_2^{(3)}=\frac{\sqrt{1-p}}{2}\sigma_y,\quad E_3^{(3)}=\frac{\sqrt{1-p}}{2}\sigma_z,
\end{aligned}
\end{eqnarray}
where $\sigma_y$ is the Pauli $Y$ operator and $p$ is the probability that no depolarization occurs. In all noisy channels ($\varepsilon_1$, $\varepsilon_2$, and $\varepsilon_3$) considered above, the parameter $p=1$ corresponds to the ideal noiseless quantum channel.

\section{Robustness of sequential Bell nonlocality sharing against channel noise}

%%%%%%%%%%%%%%%%%%%%%%%%%%%%%%%%%%%%%%%%%%%%%%%%%%%%%%%%%%%%%%%%%%%%%%%%%%%%%%%%%%%%%%%%%%%%%%%%%%%%%%%%
\begin{figure}
\epsfig{figure=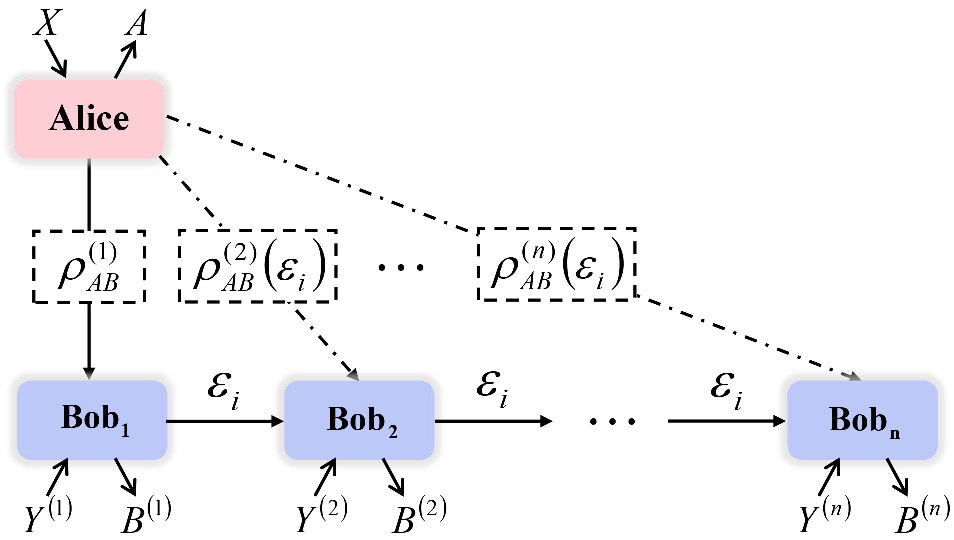,width=0.48\textwidth}
\caption{Schematic illustration of the sequential sharing of bipartite Bell nonlocality under local noisy quantum channels. An initial two-qubit state $\rho_{AB}^{(1)}$ is shared between Alice and Bob$_{1}$. Bob$_{1}$ first performs a measurement on his particle, and transmits the post-measurement qubit via a noisy quantum channel $\varepsilon_i$ to Bob$_{2}$ with no classical information about the measurement choices and outcomes. Then, Bob$_{2}$ repeats this measurement-and-sending process, and so on up to Bob$_n$. Upon performing her measurement, Alice establishes Bell-CHSH correlations with the sequence of Bobs. Let $(X,A)$ represent the binary input and output of Alice.}
\label{f1}
\end{figure}
%%%%%%%%%%%%%%%%%%%%%%%%%%%%%%%%%%%%%%%%%%%%%%%%%%%%%%%%%%%%%%%%%%%%%%%%%%%%%%%%%%%%%%%%%%%%%%%%%%%%%%%%

We consider the sequential sharing of quantum nonlocality in the CHSH scenario under the noisy quantum channels $\varepsilon_{i}$s, where a single Alice attempts to share bipartite quantum nonlocal correlations with a sequence of independent Bobs. As illustrated in Fig \ref{f1}, a two-qubit entangled state $\rho_{AB}^{(1)}$ is initially shared between Alice and Bob$_1$, let $X$ and $A$ denote the binary input and output for Alice respectively, and let $Y^{(k)}$ and $B^{(k)}$ denote the corresponding binary input and output for Bob$_{k}$ with $k\in \mathbb{N}$. In the SSQN protocol, Bob$_{1}$ performs a measurement on his particle with input $Y^{(1)}$, obtains outcome $B^{(1)}$, and then transmits the post-measurement qubit through a noisy quantum channel $\varepsilon_{i}$ to the next independent observer Bob$_2$ who repeats this process, and so on up to Bob$_n$.

In the measurement-and-sending process under local noisy channels $\varepsilon_i$s, the two-qubit state shared by Alice and Bob$_{k-1}$ is denoted by $\rho_{AB}^{(k-1)}(\varepsilon_i)$ before Bob$_{k-1}$'s local measurement. After Bob$_{k-1}$ carries out the measurement $Y^{(k-1)}=y$ and obtains the outcome $B^{(k-1)}=b$, the post-measurement state is updated according to the L\"{u}ders rule by averaging over all possible inputs and outputs of Bob$_{k-1}$ \cite{pp96lnpm}. The particle is then transmitted to Bob$_{k}$ through the noisy quantum channel $\varepsilon_i$, and the two-qubit state shared by Alice and Bob$_k$ at this stage is given by 
\begin{equation}\label{rAB}
\begin{aligned}
\rho_{AB}^{(k)}(\varepsilon_i)\!=&\!\sum\limits_{m}(\mathbb{I}\!\otimes\! E_m^{(i)})\Biggl[\frac{1}{2}\sum\limits_{b,y}(\mathbb{I}\!\otimes\!\sqrt{B^{(k-1)}_{b|y}})\\ & \rho_{AB}^{(k-1)}(\varepsilon_i)
(\mathbb{I}\otimes\sqrt{B^{(k-1)}_{b|y}})\Biggr](\mathbb{I}\otimes E_m^{(i)\dagger}),
\end{aligned}
\end{equation}
where $\sqrt{B^{(k-1)}_{b|y}}$ is the measurement effect corresponding to outcome $b$ when Bob$_{k-1}$ performs the measurement specified by input $y$, and $\{E_m^{(i)}\}$ are the Kraus operators of the noisy channel $\varepsilon_i$ with $i=1,2,3$ representing the phase-flip, bit-flip, and depolarizing quantum channels in Eqs. \eqref{phase}-\eqref{dep}. Furthermore, for Alice and Bob$_k$, the expectation value of the CHSH operator defined in Eq. \eqref{eq:CHSH} can be written as
\begin{equation}\label{Ichsh^k}
\begin{aligned}
I^{(k)}_{\mathcal{CHSH}}(\varepsilon_i)=&\tr[\rho_{AB}^{(k)}(\varepsilon_i)(X_{0}Y_{0}^{(k)}\!+\!X_{1}Y_{0}^{(k)}\!+\!X_{0}Y_{1}^{(k)}\\
&-X_{1}Y_{1}^{(k)})],
\end{aligned}
\end{equation}
where $X_i$ and $Y_j^{(k)}$ with $i,j \in \{0,1\}$ denote the local observables of Alice and Bob$_k$, respectively.

In the SSQN protocol evaluated by the CHSH inequality, the initial two-qubit state $\rho_{AB}^{(1)}=\proj{\psi}$ is chosen to be a maximally entangled Bell state
\begin{equation}\label{bell-state}
\ket{\psi}=\frac{1}{\sqrt{2}}(\ket{00}+\ket{11}).
\end{equation}
The weak measurement strategy is described by two-outcome positive operator-valued measurements (POVMs) consisting of $\{M, \mathbb{I}-M\}$. Moreover, the measurement operator can be written as $M=\frac{1}{2}(\mathbb{I}+\gamma\sigma_{\vec{r}})$, where the parameter $\gamma\in[0,1]$ quantifies the measurement sharpness with $\gamma=1$ corresponding to a projective (sharp) measurement and $\sigma_{\vec{r}}=r_x\sigma_x+r_y\sigma_y+r_z\sigma_z$ is a generic Pauli operator with $\vec{r}=(r_x,r_y,r_z)\in\mathbb{R}$ satisfying $||\vec{r}||=1$. For Alice, the measurements $X_i$ with $i=0,1$ are described by two-outcome POVMs with elements $\{A_{0|i}, A_{1|i}\}$, where $ A_{1|i}=\mathbb{I}-A_{0|i}$. Similarly, for each observer Bob$_k$ ($k\in \mathbb{N}$), the measurements $Y_j^{(k)}$ with $j=0, 1$ are defined by two operators $\{B_{0|j}^{(k)}, B_{1|j}^{(k)}\}$ satisfying $B_{1|j}^{(k)}=\mathbb{I}-B_{0|j}^{(k)}$.

To analyze the robustness of sequential Bell nonlocality sharing against noise in quantum channels, we consider two kinds of measurement strategies. The first strategy, referred to as MS-I, was proposed by Brown and Colbeck in Ref. \cite{pjb20prl} for the noiseless unbounded sequential sharing of bipartite Bell-CHSH nonlocality. The POVM operators for Alice in strategy MS-I take the form
\begin{equation}\label{bellA}
\begin{aligned}
A_{0|0}=\frac{1}{2}[\mathbb{I}+\cos(\theta)\sigma_z+\sin(\theta)\sigma_x],\\
A_{0|1}=\frac{1}{2}[\mathbb{I}+\cos(\theta)\sigma_z-\sin(\theta)\sigma_x],
\end{aligned}
\end{equation}
where the parameter $\theta\in(0,\pi/4]$. The POVM operators for the independent observer Bob$_k$ are expressed as
\begin{eqnarray}\label{bellB}
B_{0|0}^{(k)}=\frac{1}{2}(\mathbb{I}+\sigma_z),
\quad B_{0|1}^{(k)}=\frac{1}{2}(\mathbb{I}+\gamma_k\sigma_x),
\end{eqnarray}
where $\gamma_k\in[0,1]$ denotes the measurement sharpness and $k\in\mathbb{N}$ for any given number $N$ of Bobs in the sequence. The second measurement strategy, which we refer to as MS-II, is introduced here. The associated POVMs for Alice and Bob$_{k}$ in strategy MS-II are given by
\begin{equation}\label{bellAB}
\begin{aligned}
&A_{0|0}=\frac{1}{2}[\mathbb{I}+\cos(\theta)\sigma_x+\sin(\theta)\sigma_z],\\
&A_{0|1}=\frac{1}{2}[\mathbb{I}+\cos(\theta)\sigma_x-\sin(\theta)\sigma_z],\\
&B_{0|0}^{(k)}=\frac{1}{2}(\mathbb{I}+\sigma_x),\quad B_{0|1}^{(k)}=\frac{1}{2}(\mathbb{I}+\gamma_k\sigma_z),
\end{aligned}
\end{equation}
where the parameters are $\theta\in(0,\pi/4]$, $\gamma_k\in[0,1]$, and $k\in\mathbb{N}$. It can be proven that strategy MS-II is able to accomplish the unbounded SSQN in the noiseless scenario as well (see Appendix A).

Next, we study the noise robustness of sequential Bell-CHSH nonlocality sharing under the strategies MS-I and MS-II, assuming that Alice and Bob$_1$ initially share the maximally entangled two-qubit Bell state given in Eq. \eqref{bell-state}. As shown in Fig. \ref{f1}, we first consider that Alice and the sequence of Bobs employ strategy MS-I given in Eqs. \eqref{bellA} and \eqref{bellB}, and that the local channel is subject to phase-flip noise described by $\varepsilon_1$ in Eq. \eqref{phase}. Combining Eqs. \eqref{rAB}-\eqref{bell-state}, we can obtain that the expected CHSH value between Alice and Bob$_k$ for strategy MS-I on the two-qubit state $\rho_{AB}^{(k)}(\varepsilon_1)$ takes the form
\begin{eqnarray}\label{bellCHSH}
	\begin{aligned}
		I_{\mathcal{CHSH}}^{(k)}(\varepsilon_1)_{\mathrm{MS\text{-}I}}=&2^{2-k}\Biggl[\gamma_k \sin(\theta)(2p-1)^{k-1} \\ &\!+\!\cos(\theta)\prod_{j=1}^{k-1}\left(1\!+\!\sqrt{1-\gamma_j^{2}}\right)\Biggr],
	\end{aligned}
\end{eqnarray}
where $p$ is the channel parameter representing the probability that the qubit remains unchanged, and the case of $p=1$ reproduces the noiseless CHSH expectation value derived in Ref. \cite{pjb20prl}. Without loss of generality, the noise parameter in Eq. \eqref{bellCHSH} is set to $p>1/2$, since we can obtain the same expression via an equivalent initial Bell state for the case of $p<1/2$. It is noted that, when the noise parameter is $p=1/2$ for quantum channels, the output state is a separable state. To ensure the sharing of quantum nonlocality, the expected CHSH value must satisfy $I_{\mathcal{CHSH}}^{(k)}(\varepsilon_1)_{\mathrm{MS\text{-}I}}>2$, which further imposes a constraint on the value of $\gamma_k$ determined by Eq. \eqref{bellCHSH}, leading to the requirement
\begin{eqnarray}\label{16}
\gamma_k>\frac{2^{k-1}-\cos(\theta)\prod_{j=1}^{k-1}(1+\sqrt{1-\gamma_j^{2}})}{\sin(\theta)(2p-1)^{k-1}}.
\end{eqnarray}
Consequently, we can define a new sequence $\{\gamma_k(\theta)\}_{k\in \mathbb{N}}$ by introducing a parameter $\epsilon>0$, which is given by
\begin{equation}\label{17}
	\gamma_{k}(\theta)\!=\!\left\{
	\begin{array}{ll}
		(1\!+\!\epsilon)\frac{1-\cos(\theta)}{\sin(\theta)}, & \hbox{$k\!=\!1$} \\
		(1\!+\!\epsilon)\frac{2^{k-1}-\cos(\theta)P_k}{\sin(\theta)(2p-1)^{k-1}} , & \hbox{$0\!\leq \!\gamma_{k-1}(\theta)\!\leq \!1$ } \\
		\infty, & \hbox{otherwise}
	\end{array},
	\right.
\end{equation}
where $P_{k}=\prod_{j=1}^{k-1}(1+\sqrt{1-\gamma_j^{2}})$, the parameter $\theta\in (0,\theta_k)$ with $\theta_{k}\in(0,\pi/4]$ for each $0<\gamma_{k}(\theta)<1$, and $\gamma_k=\infty$ indicating that the violation of the CHSH inequality is impossible for Alice and Bob$_k$. For the local bit-flip channel $\varepsilon_2$ and the depolarizing channel $\varepsilon_3$, we can make a similar analysis based on Eqs. \eqref{rAB}-\eqref{bell-state}, examining both the expected CHSH value and the requirement for the sharpness parameters. Furthermore, this treatment can be extended to the measurement strategy MS-II given in Eq. \eqref{bellAB}, considering the same three local noisy channels $\varepsilon_{i}$s with $i=1, 2, 3$. We note that, for the bit-flip channel $\varepsilon_2$, the expected CHSH value $I_{\mathcal{CHSH}}^{(k)}(\varepsilon_2)_{\mathrm{MS\text{-}II}}$ for strategy MS-II on the two-qubit state $\rho_{AB}^{(k)}(\varepsilon_2)$ has the same expression as that in Eq. \eqref{bellCHSH}. Then, by analyzing the properties of the sequence of sharpness parameters $\{\gamma_k(\theta)\}$ for the two measurement strategies MS-I and MS-II in the presence of the three local noisy channels, we arrive at the following theorem.

%%%%%%%%%%%%%%%%%%%%%%%%%%%%%%%%%%%%%%%%%%%%%%%%%%%%%%%%%%%%%%%%%%%%%%%%%%%%%%%%%%%%%%%%%%%%%%%%%%%%%%%%%
\textit{Theorem 1.}---For the phase-flip quantum channel $\varepsilon_1$ with $p>1/2$, there exists, for each $n\in \mathbb{N}$, a sequence $\{\gamma_{k}(\theta)\}_{k=1}^{n}$ and a $\theta_{n}\in(0,\pi/4]$ under strategy MS-I for the initial Bell state such that $I^{(k)}_{\mathcal{CHSH}}(\varepsilon_1)_{\mathrm{MS\text{-}I}}>2$ for $k=1,2,\cdots, n$. This unbounded sequential sharing can extend to the bit-flip channel $\varepsilon_2$ ($p>1/2$) by switching to strategy MS-II, enabling arbitrarily many independent Bobs to share
Bell-CHSH nonlocality of the initial state with a single Alice, yet is unattainable under local depolarizing channel $\varepsilon_3$ for either measurement strategy.
%%%%%%%%%%%%%%%%%%%%%%%%%%%%%%%%%%%%%%%%%%%%%%%%%%%%%%%%%%%%%%%%%%%%%%%%%%%%%%%%%%%%%%%%%%%%%%%%%%%%%%%%%

The proof of Theorem 1 is provided in Appendix A, where it is shown that, for local channels with the noise of phase-flip $\varepsilon_1$ or bit-flip $\varepsilon_2$, a single Alice can sequentially share bipartite quantum nonlocality of the initial Bell state with arbitrarily many independent Bobs by employing the corresponding measurement strategies (MS-I/MS-II) associated with a sequence of sharpness $\{\gamma_{k}(\theta)\}_{k=1}^{n}$ and a properly chosen angle parameter $\theta$. Therefore, since the unbounded SSQN protocol is robust against phase-flip and bit-flip noise, we refer to $\varepsilon_1$ as the noise-immune channel under MS-I and to $\varepsilon_2$ as the noise-immune one under MS-II, respectively. This strategy‑dependent noise immunity underpins the unbounded sequential sharing of bipartite quantum nonlocality of the initial Bell state. In contrast, for other noisy channels under the two measurement strategies, the unbounded SSQN is destroyed and cannot be realized.

\begin{figure}
	\epsfig{figure=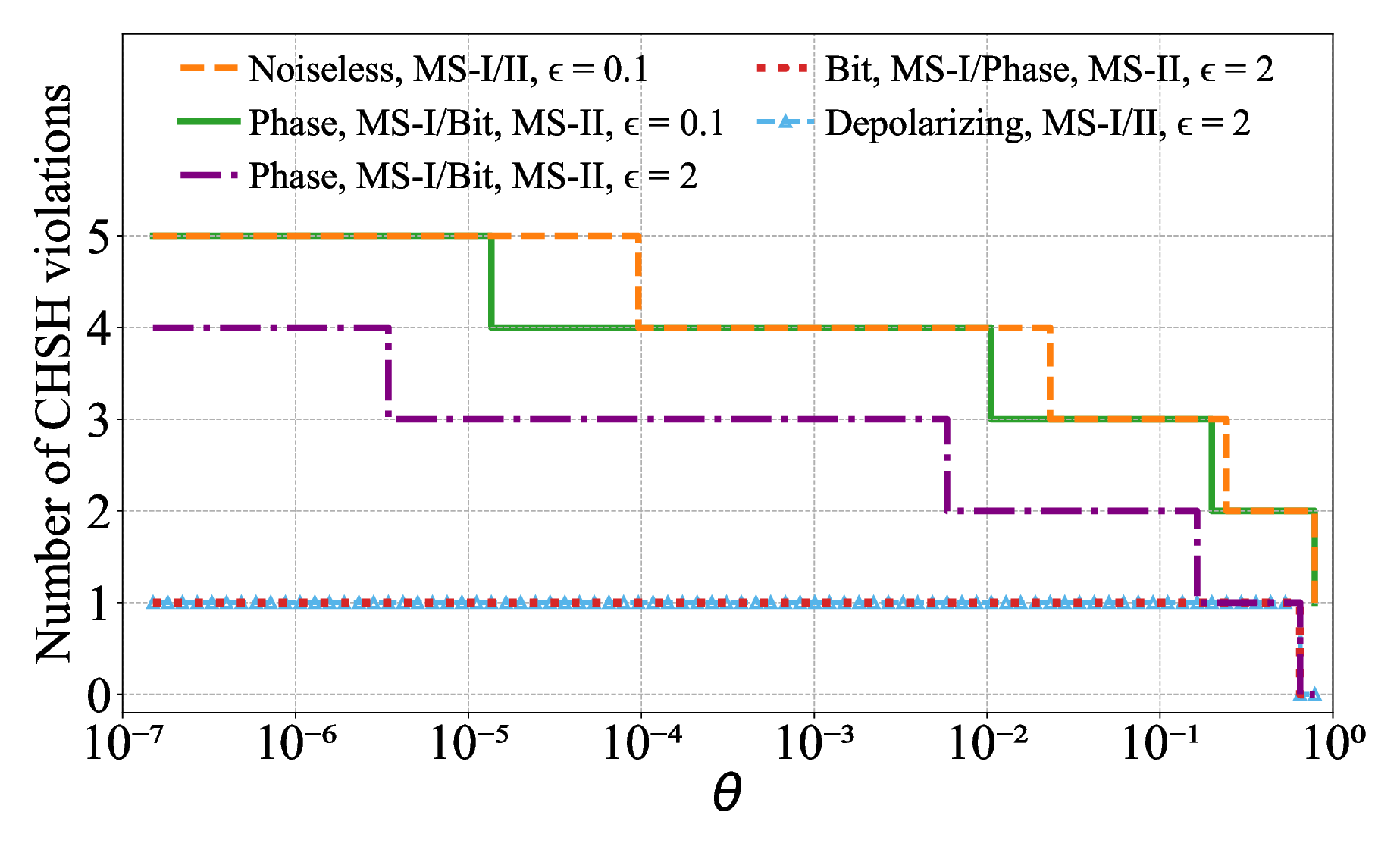,width=0.48\textwidth}
	\caption{The number of Bobs that can violate the CHSH inequality as a function of $\theta$ for the strategies MS-I and MS-II, under noiseless and noisy quantum channels. Alice and Bob$_1$ initially share a Bell state $\ket{\psi}$ and the noise parameter is set to $p=0.95$ for phase-flip, bit-flip and depolarizing channels. By employing the suitable measurement strategy associated with the appropriate values for $\theta$ and $\epsilon$, the SSQN protocol is completely robust against phase-flip and bit-flip noise, while the sequential sharing is destroyed by depolarizing noise in local channels.}
	\label{f2}
\end{figure}

To further illustrate the impact of noise in quantum channels, we examine a specific example involving a single Alice and a sequence of five Bobs, and analyze the factors that limit the number of sequential nonlocality sharings. For a given channel noise, the number of Bobs who can sequentially share the nonlocality is determined by the parameters $\theta$ and $\epsilon$ in the strategies MS-I and MS-II. As shown in Fig. \ref{f2}, we plot the number of Bobs that can violate CHSH inequality as a function of the parameter $\theta$ (on a logarithmic scale) for two fixed values of $\epsilon$ under different measurement strategies, where both the noiseless and noisy channels are considered and we set the noise parameter to $p=0.95$ for all noisy channels $\varepsilon_1$, $\varepsilon_2$ and $\varepsilon_3$. The orange dashed line corresponds to the results of the noiseless SSQN protocol under the strategies MS-I and MS-II, where the maximal number $n=5$ for Bobs that sequentially share the Bell nonlocality can be achieved by adjusting the parameter $\theta$ for the fixed value $\epsilon=0.1$. The green solid line shows the results for the noise-immune channels (the phase-flip channel under strategy MS-I and the bit-flip channel under strategy MS-II with the same parameter $\epsilon=0.1$), where the maximal number $n=5$ for Bobs is realized by appropriately decreasing the parameter $\theta$ relative to the noiseless case. Moreover, when the parameter $\epsilon$ is increased to $\epsilon=2$, the SSQN protocol achieves at most $n=4$ (the purple dot-dashed line) for $\theta\geq 1.5\times 10^{-7}$. These results indicate that the complete robustness of the SSQN protocol against phase-flip and bit-flip noise requires the joint adjustment of the parameters $\theta$ and $\epsilon$. In addition, for non-immune channels such as the bit-flip under MS-I, phase-flip under MS-II, and depolarizing channels, the unbounded SSQN of the Bell state is destroyed, as shown by the red-dotted line and blue dashed line with triangle markers in Fig. \ref{f2}.

%%%%%%%%%%%%%%%%%%%%%%%%%%%%%%%%%%%%%%%%%%%%%%%%%%%%%%%%%%%%%%%%%%%%%%%%%%%%%%%%%%%%%%%%%%%%%%%%%%%%%%%%%
%%%%%%%%%%%%%%%%%%%%%%%%%%%%%%%%%%%%%%%%%%%%%%%%%%%%%%%%%%%%%%%%%%%%%%%%%%%%%%%%%%%%%%%%%%%%%%%%%%%%%%%%%

\section{Robustness of sequential Mermin nonlocality sharing against channel noise}

\begin{figure}
	\epsfig{figure=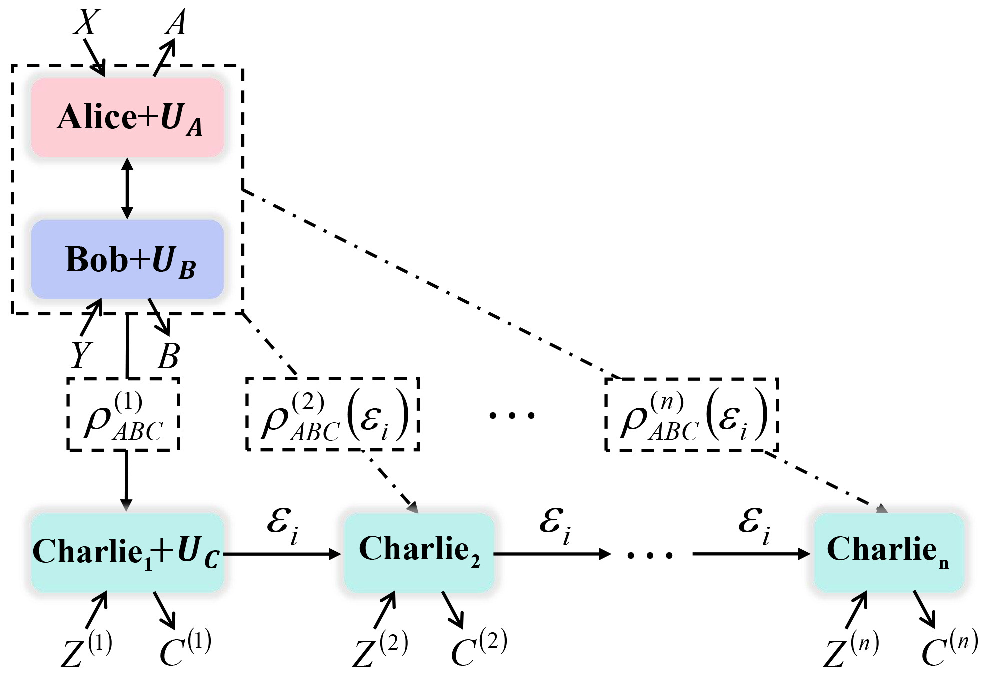,width=0.48\textwidth}
	\caption{Schematic of the SSQN in tripartite Mermin scenario under noisy quantum channels. The initial three-qubit state $\rho_{ABC}^{(1)}$ is distributed among Alice, Bob, and Charlie$_{1}$, which is adaptively modulated by the local unitary operations $U_A$, $U_B$ and $U_C$ determined by different measurement strategies. After Charlie$_{1}$ performs a randomly selected measurement $Z^{(1)}$ and records the outcome $C^{(1)}$, he transmits the resulting post-measurement qubit to Charlie$_{2}$ through a noisy quantum channel $\varepsilon_i$. The same process is applied iteratively until the observer Charlie$_n$ is reached, where the measurement choice and the outcome of each Charlie are kept independent throughout the sequence. Here $(X,A)$ and $(Y,B)$ denote the binary input and output for Alice and Bob.}
	\label{f3}
\end{figure}

In this section, we consider a tripartite scenario in which a single Alice and a single Bob try to sequentially share Mermin nonlocality with a sequence of independent Charlies, under the influence of local noisy channels $\varepsilon_i$s (phase-flip, bit-flip, and depolarizing noise). As illustrated in Fig.~\ref{f3}, Alice, Bob, and Charlie$_1$ initially share a three-qubit entangled state $\rho_{ABC}^{(1)}$, which is adaptively modulated by appropriate local unitary operations $U_A$, $U_B$ and $U_C$ performed by the three parties with their forms determined by different measurement strategies. Let $X$ and $A$ denote Alice's binary input and output, respectively, and $Y$ and $B$ for Bob. Moreover, for each $k \in \mathbb{N}$, let $Z^{(k)}$ and $C^{(k)}$ represent the binary input and output of Charlie$_k$. In the protocol, Charlie$_{1}$ performs a measurement  $Z^{(1)}$ on his particle and obtains outcome $C^{(1)}$. Subsequently, this post-measurement qubit is transmitted by Charlie$_1$ through a noisy quantum channel $\varepsilon_{i}$ to the next independent Charlie$_2$, who repeats the measurement-sending operation. The process continues iteratively until the final observer Charlie$_n$.

%%%%%%%%%%%%%%%%%%%%%%%%%%%%%%%%%%%%%%%%%%%%%%%%%%%%%%%%%%%%%%%%%%%%%%%%%%%%%%%%%%%%%%%%%%%%%%%%%%%%%%%%%
%%%%%%%%%%%%%%%%%%%%%%%%%%%%%%%%%%%%%%%%%%%%%%%%%%%%%%%%%%%%%%%%%%%%%%%%%%%%%%%%%%%%%%%%%%%%%%%%%%%%%%%%%

Under the local noisy channel $\varepsilon_i$, we denote by $\rho_{ABC}^{(k-1)}(\varepsilon_i)$ the quantum state shared among Alice, Bob and Charlie$_{k-1}$ prior to Charlie$_{k-1}$'s local measurement. After Charlie$_{k-1}$ performs the measurement $Z^{(k-1)}=z$ and obtains the outcome $C^{(k-1)}=c$, the post-measurement particle is then transmitted to Charlie$_{k}$ via the local noisy channel $\varepsilon_i$. At this stage, the tripartite quantum state shared by Alice, Bob and Charlie$_{k}$ can be written as 
\begin{equation}\label{rABC}
	\begin{aligned}
		\rho_{ABC}^{(k)}(\varepsilon_i)\!\!=\!\!\!&\sum\limits_{m}(\mathbb{I}\!\otimes\!\mathbb{I}
		\!\otimes \!E_m^{(i)})\!\Biggl[\!\frac{1}{2}\!\sum\limits_{c,z}(\mathbb{I}\!\otimes\!\mathbb{I}
		\!\otimes\!\sqrt{C^{(k-1)}_{c|z}}) \\ &
		\rho_{ABC}^{(k-1)}(\varepsilon_i)\!(\mathbb{I}\!\otimes\!\mathbb{I}\!\otimes\!\sqrt{C^{(k-1)}_{c|z}})\Biggr]\!(\mathbb{I}\!\otimes\!\mathbb{I}
		\!\otimes\! E_m^{(i)\dagger}),
	\end{aligned}
\end{equation}
where $\sqrt{C^{(k-1)}_{c|z}}$ denotes the POVM effect corresponding to the outcome $c$ of Charlie$_{k-1}$'s measurement with input $z$, and $\{E_m^{(i)}\}$ are the Kraus operators of the noisy channels $\varepsilon_i$s defined in Eqs. \eqref{phase}-\eqref{dep}. Consequently, the expectation value of the tripartite Mermin operator defined in Eq. \eqref{eq:Mermin} for Alice, Bob and Charlie$_{k}$ is given by 
\begin{equation}\label{Ichsh^k}
\begin{aligned}
I^{(k)}_{\mathcal{M}}(\varepsilon_i)=\tr[\rho_{ABC}^{(k)}(\varepsilon_i)&(X_{1}Y_{0}Z_{0}^{(k)}+ X_{0}Y_{1}Z_{0}^{(k)}\\& +X_{0}Y_{0}Z_{1}^{(k)}-X_{1}Y_{1}Z_{1}^{(k)})],
\end{aligned}
\end{equation}
where $X_j$, $Y_j$ and $Z_j^{(k)}$ with $j \in \{0,1\}$ are the local measurements performed by the three observers. Alice's two measurement settings are labeled by $X_j$  $(j = 0, 1)$, described by POVM operators $\{A_{0|j}, A_{1|j}\}$ satisfying $A_{1|j} = \mathbb{I} - A_{0|j}$. Similarly, Bob's two measurement inputs $Y_j $s are described by the POVM operators $\{B_{0|j}, B_{1|j}\}$ with $B_{1|j} = \mathbb{I} - B_{0|j}$ for $j=0, 1$. For each independent Charlie$_k$ with $k \in \mathbb{N}$, the two measurement inputs $Z_j^{(k)}$ $(j = 0, 1)$ are implemented via $\{C_{0|j}^{(k)}, C_{1|j}^{(k)}\}$ with $C_{1|j}^{(k)} = \mathbb{I} - C_{0|j}^{(k)}$.

\subsection{Noisy scenario for SSQN with GHZ states}

In the presence of local noisy channels, we consider the SSQN evaluated by the violation of tripartite Mermin inequality, where the initial state $\rho_{ABC}^{(1)}=\proj{\psi}$ shared by Alice, Bob and Charlie$_1$ is a three-qubit GHZ state
\begin{equation}\label{GHZ}
	|\psi\rangle_{GHZ}=\frac{1}{\sqrt{2}}(|000\rangle+|111\rangle).
\end{equation}
To investigate the robustness of sequential Mermin nonlocality sharing against local noise in quantum channels, we consider two kinds of measurement strategies, referred to as MS-III and MS-IV, respectively.

The strategy MS-III was originally introduced by Xi \textit{et al} in Ref. \cite{yx23pra} for the noiseless case, where it was shown to enable the sequential sharing of tripartite Mermin nonlocality among Alice, Bob, and an arbitrary number of independent Charlies. In this strategy, the POVM operators for Alice's two measurements are
\begin{eqnarray}\label{GHZA}
	A_{0|0}=\frac{\mathbb{I}+\sigma_x}{2},\quad A_{0|1}=\frac{\mathbb{I}+\sigma_y}{2},
\end{eqnarray}
the POVM operators for Bob are given by
\begin{eqnarray}\label{GHZB}
	B_{0|0}=\frac{\mathbb{I}-\theta\sigma_y}{2},\quad B_{0|1}=\frac{\mathbb{I}+\theta\sigma_x}{2},
\end{eqnarray}
where $\theta\in(0,1)$ is the sharpness parameter, and the POVM operators associated with Charlie$_{k}$'s measurements are 
\begin{eqnarray}\label{GHZC}
	C_{0|0}^{(k)}=\frac{\mathbb{I}+\sigma_x}{2},\quad C_{0|1}^{(k)}=\frac{\mathbb{I}+\gamma_k\sigma_y}{2},
\end{eqnarray}
where the parameter $\gamma_k\in[0,1]$ for each $k=1$,\ldots,$n$ with any given number $n\in \mathbb{N}$.

We propose a new strategy MS-IV assisted by local unitary operations on the initially shared GHZ state, which can be proven to achieve the unbounded SSQN evaluated by the violation of Mermin inequality in the noiseless scenario (see Appendix B for the proof). In this strategy, the local unitary operation $U_L$ is performed by Alice, Bob and Charlie$_1$, which takes the form
\begin{eqnarray}\label{U}
	U_L=U_A\otimes U_B\otimes U_{C_1},
\end{eqnarray}
where each local operation consists of a Hadamard gate
\begin{equation}
    U_i = \frac{1}{\sqrt{2}} \begin{pmatrix} 1 & 1 \\ 1 & -1 \end{pmatrix}
\end{equation}
where $i=A, B$, and $C_1$, respectively. Upon applying the local unitary $U_L$, the resulting quantum state is
\begin{equation}
	\begin{aligned}
		|\psi\rangle_{GHZ'}&=\frac{1}{\sqrt{2}}(\vert \tilde{0}\tilde{0}\tilde{0}\rangle+\vert \tilde{1}\tilde{1}\tilde{1}\rangle)\\
		   &=\frac{1}{2} (\vert 000\rangle + \vert 011\rangle + \vert 101\rangle + \vert 110\rangle)
	\end{aligned}
\end{equation}
where the local basis is rotated to $\tilde{\vert0\rangle}=(\vert0\rangle+\vert1\rangle)/\sqrt{2}$ and $\tilde{\vert1\rangle}=(\vert0\rangle-\vert1\rangle)/\sqrt{2}$, transforming the state into the so-called tetrahedral state which plays an important role in quantum teleportation \cite{pa06pra,ab19pra}. In strategy MS-IV, the POVM operators corresponding to the measurements performed by Alice, Bob, and Charlie$_{k}$ can be expressed as
\begin{eqnarray}\label{27}
	\begin{aligned}
	&A_{0|0}=\frac{\mathbb{I}+\sigma_z}{2},\quad A_{0|1}=\frac{\mathbb{I}-\sigma_y}{2},\\
	&B_{0|0}=\frac{\mathbb{I}+\theta\sigma_y}{2},\quad B_{0|1}=\frac{\mathbb{I}+\theta\sigma_z}{2},\\
	&C_{0|0}^{(k)}=\frac{\mathbb{I}+\sigma_z}{2},\quad C_{0|1}^{(k)}=\frac{\mathbb{I}-\gamma_k\sigma_y}{2},	
	\end{aligned}
\end{eqnarray}
where $\theta\in(0,1)$ and $\gamma_k\in[0,1]$ are sharpness parameters, and $k$ runs from $1$ to $n$ for any given $n\in\mathbb{N}$. 

We now proceed to analyze the noise robustness of sequential Mermin nonlocality sharing under different measurement strategies, where Alice, Bob, and Charlie$_{1}$ initially share the three-qubit GHZ state in Eq. \eqref{GHZ}. As shown in Fig. \ref{f3}, we first consider the case where local quantum channels are subject to bit-flip noise described by $\varepsilon_2$ in Eq. \eqref{bit}, and Alice, Bob, and the sequence of Charlies adopt strategy MS-III defined in Eqs. \eqref{GHZA}-\eqref{GHZC}, which does not involve local unitary operations and corresponds to the case $U_A=U_B=U_C=I$. Combining Eqs. \eqref{rABC}-\eqref{GHZ}, we find that the expected Mermin value for the three observers Alice, Bob and Charlie$_k$ under the strategy MS-III can be expressed as
\begin{equation}\label{IM b G}
	I^{(k)}_{\mathcal{M}}\!(\!\varepsilon_2\!)_{\mathrm{MS\text{-}III}}\!\!=\! \!2\theta\!\! \left[\!\gamma_k(p\!-\!\frac{1}{2})^{k\!-\!1} \!\!+\!2^{1\!-\!k}\!\prod_{j=1}^{k\!-\!1}\!(\!1\!\!+\!\!\sqrt{\!1\!\!-\!\gamma_j^{2}})\!\right]\!\!,
\end{equation}
where $\theta$ and $\gamma_k$ are sharpness parameters, and the channel parameter is set to $p>1/2$. In order to share the Mermin nonlocality, the condition $I_{\mathcal{M}}^{(k)}(\varepsilon_2)_{\mathrm{MS\text{-}III}}>2$ must be satisfied, which leads to the relation
\begin{equation}\label{29}
	\gamma_{k}>(p-\frac{1}{2})^{1-k}[\frac{1}{\theta}-2^{1-k}\prod\limits_{j=1}^{k-1}(1+\sqrt{1-\gamma_{j}^{2}})].
\end{equation}
Furthermore, let $\epsilon>0$, we can define ${\gamma_k(\theta)}$ as
\begin{equation}\label{30}
	\gamma_{k}(\theta)=\left\{
	\begin{array}{ll}
		(1+\epsilon)(\frac{1}{\theta}-1), & \hbox{$k=1$,} \\[2mm]
		(1+\epsilon)\frac{\frac{1}{\theta}-2^{1-k}P_{k}}{(p-\frac{1}{2})^{k-1}} , & \hbox{$0\leq\gamma_{k-1}(\theta)\leq1$ ,} \\[2mm]
		\infty, & \hbox{otherwise,}
	\end{array}
	\right.
\end{equation}
where $P_{k}=\prod_{j=1}^{k-1}(1+\sqrt{1-\gamma_j^{2}})$, the parameter $\theta\in (\theta_k,1)$ with $\theta_{k}\in(0,1)$ for all $0<\gamma_{k}(\theta)<1$. For the noisy SSQN under the phase-flip channel $\varepsilon_1$ and the depolarization channel $\varepsilon_3$ using strategy MS-III, we analyze the expected Mermin values $I^{(k)}_{\mathcal{M}}$s and the corresponding sharpness parameter $\gamma_{k}(\theta)$s in a similar manner. Moreover, by adopting our newly proposed measurement strategy MS-IV assisted by local unitary operations $U_L$ applied to the initial shared GHZ state, we can derive the corresponding expected Mermin values in the presence of the three local noisy channels $\varepsilon_i$s given in Eqs. \eqref{phase}-\eqref{dep}. It is worth noting that, for the phase-flip channel $\varepsilon_1$ with the noise parameter $p>1/2$, the expected Mermin value $I^{(k)}_{\mathcal{M}}(\varepsilon_1)_{\mathrm{MS\text{-}IV}}$ under strategy MS-IV takes the same form as that in Eq. \eqref{IM b G}. Detailed derivations of these expected values for different noisy channels and measurement strategies are presented in Appendix B. Then, by analyzing the properties of the sequences $\{\gamma_{k}(\theta)\}$ for the strategies MS-III and MS-IV in the presence of these local noisy channels, we obtain the following theorem. 

%%%%%%%%%%%%%%%%%%%%%%%%%%%%%%%%%%%%%%%%%%%%%%%%%%%%%%%%%%%%%%%%%%%%%%%%%%%%%%%%%
\textit{Theorem 2.}---Starting from an initial GHZ state, for the bit-flip channel $\varepsilon_2$ with $p>1/2$, there exists, for each $n\in\mathbb{N}$, a sequence $\{\gamma_k(\theta)\}_{k=1}^n$ and a $\theta_n\in(0,1)$ under strategy MS--III such that $I^{(k)}_{\mathcal{M}}(\varepsilon_2)_{\mathrm{MS\text{-}III}}>2$ for all $k\le n$. This unbounded sequential sharing of Mermin nonlocality persists against the phase-flip channel $\varepsilon_1$ with $p>1/2$ via strategy MS-IV assisted by the local unitary $U_L$ on the initial state, but is destroyed by local depolarizing channel $\varepsilon_3$ under either strategy.
%%%%%%%%%%%%%%%%%%%%%%%%%%%%%%%%%%%%%%%%%%%%%%%%%%%%%%%%%%%%%%%%%%%%%%%%%%%%%%%%%

A detailed proof of Theorem 2 is provided in Appendix B, which shows that an arbitrary number of Charlies can sequentially share Mermin nonlocality with Alice and Bob under noise-immune quantum channels, provided a suitable measurement strategy is employed. Strategy MS-III renders the SSQN protocol immune to the bit-flip channel $\varepsilon_2$, whereas strategy MS-IV assisted by local unitary operations provides immunity to phase-flip noise in the quantum channel $\varepsilon_1$. Depending on the type of channel noise described by $\varepsilon_1$ and $\varepsilon_2$, the observers can switch measurement strategies to achieve the unbounded SSQN evaluated by the violation of the Mermin inequality. In contrast, for all other noisy channels considered with strategies MS-III and MS-IV, the unbounded property of the SSQN based on the GHZ state is lost. When the noise parameter $p<1/2$ for quantum channels $\varepsilon_1$ and $\varepsilon_2$, the unbounded SSQN is still available via an equivalent initial state, since the same expression as that in Eq. \eqref{IM b G} can be obtained. Note that the unbounded SSQN is destroyed when the two noisy quantum channels are set to $p=1/2$, due to the output state being a separable state.

%%%%%%%%%%%%%%%%%%%%%%%%%%%%%%%%%%%%%%%%%%%%%%%%%%%%%%%%%%%%%%%%%%%%%%%%%%%%%%%%%%%%%%%%%%%%%%%%%%%%%%%%
\begin{figure}
	\epsfig{figure=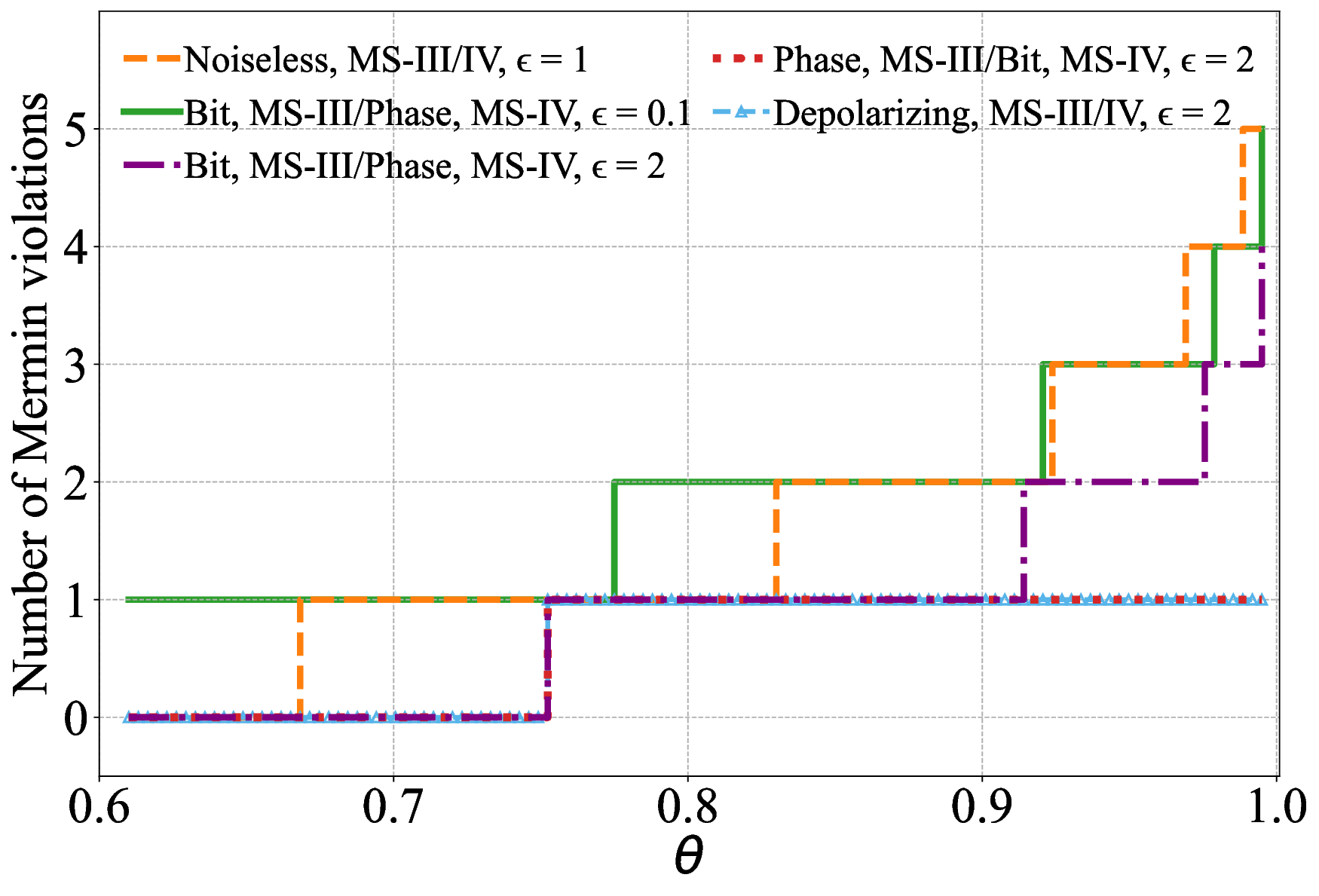,width=0.48\textwidth}
	\caption{The number of Charlies that can violate Mermin inequality as a function of $\theta$ for strategies MS-III/MS-IV under the noiseless and noisy quantum channels, where Alice, Bob and Charlie$_1$ initially share a GHZ state and the noise parameter is set to $p=0.85$ for phase-flip, bit-flip and depolarizing channels. By switching between strategies MS-III and MS-IV  and adjusting $\theta$ and $\epsilon$ appropriately, the SSQN protocol is fully robust against phase-flip and bit-flip noise, whereas the sequential sharing property is destroyed by depolarizing noise in local channels.}
	\label{f4}
\end{figure}
%%%%%%%%%%%%%%%%%%%%%%%%%%%%%%%%%%%%%%%%%%%%%%%%%%%%%%%%%%%%%%%%%%%%%%%%%%%%%%%%%%%%%%%%%%%%%%%%%%%%%%%%

Next, we examine how channel noise affects the sequential sharing of nonlocality by investigating a specific example involving a single Alice, a single Bob, and a sequence of five Charlies (\textit{i.e.}, $n=5$). In this case, the maximal number of Charlies achievable in the SSQN protocol depends on the property of noise as well as the sharpness parameter $\theta$ and the quantity $\epsilon$ in the measurement strategies. As illustrated in Fig. \ref{f4}, we plot the number of Charlies that can detect the nonlocality as a function of the parameter $\theta$ for different values of $\epsilon$, where both noiseless and noisy quantum channels are considered and the noise parameter is set to $p=0.85$ for all noisy channels $\varepsilon_1$, $\varepsilon_2$ and $\varepsilon_3$. The orange dashed line is the result for noiseless channels using strategies MS-III and MS-IV, where the maximum number $n=5$ for Charlies that sequentially share Mermin nonlocality with Alice and Bob can be realized by adjusting the parameter $\theta$ at $\epsilon=1$. The green solid line corresponds to the results of two noise-immune channels (the bit-flip channel under strategy MS-III and the phase-flip one under strategy MS-IV), where the maximal number $n=5$ for the sequence of Charlies can be achieved by setting $\epsilon=0.1$ and appropriately increasing the value of $\theta$ relative to that in the noiseless case. Moreover, when the parameter $\epsilon$ increases to $2$, the SSQN under the two noise-immune channels can reach at most $n=4$ for $\theta\leq 0.995$ (the purple dot-dashed line). In addition, for other noisy quantum channels with different measurement strategies (the red-dotted line and the blue-dashed line with triangle markers), the unbounded SSQN with the GHZ state is destroyed.

\subsection{Noisy scenario for SSQN with W states}

Under the influence of local noisy channels, we analyze the robustness of sequential Mermin nonlocality sharing using a three-qubit W state
\begin{equation}\label{W}
	|\psi\rangle_W=\frac{1}{\sqrt{3}}(|100\rangle+|010\rangle+|001\rangle).
\end{equation}
In this case, the initially shared three-qubit state in Fig. \ref{f3} is given by $\rho_{ABC}^{(1)}=|\psi\rangle_{W}\langle\psi|$. Two measurement strategies are employed to study the robustness of SSQN based on the W state against noise in local quantum channels. The first strategy, referred to as MS-V, is proposed by Shen and Li in Ref. \cite{bz24pra} and can realize the unbounded SSQN protocol using the W state in the noiseless scenario. The POVM operators for Alice's measurements have the forms
\begin{equation}\label{WA}
\begin{aligned}
	A_{0|0}=\frac{\mathbb{I}+\sin(\theta)\sigma_x+\cos(\theta)\sigma_z}{2},\\ A_{0|1}=\frac{\mathbb{I}+\sin(\theta)\sigma_x-\cos(\theta)\sigma_z}{2},
\end{aligned}
\end{equation}
and the POVM operators for Bob can be expressed as
\begin{equation}\label{WB}
\begin{aligned}
	B_{0|0}=\frac{\mathbb{I}+\sin(\theta)\sigma_x+\cos(\theta)\sigma_z}{2},\\ B_{0|1}=\frac{\mathbb{I}+\sin(\theta)\sigma_x-\cos(\theta)\sigma_z}{2},
\end{aligned}
\end{equation}
where the measurement parameter $\theta$ for Alice and Bob lies in the interval $[0,\frac{\pi}{2}]$. Moreover, the POVMs performed by Charlie$_{k}$ can be written as 
\begin{eqnarray}\label{WC}
	C_{0|0}^{(k)}=\frac{\mathbb{I}+\sigma_z}{2},\quad C_{0|1}^{(k)}=\frac{\mathbb{I}+\gamma_k\sigma_x}{2},
\end{eqnarray}
where the sharpness parameter satisfies $\gamma_k\in[0,1]$, and the index of Charlie is $k=1, \dots, n$ for any given $n\in\mathbb{N}$. We also propose another strategy, referred to as MS-VI, which is assisted by local unitary operations on the initially shared W state. In this case, Alice, Bob, and Charlie$_1$ apply the local unitary operation $U_L$ given in Eq. \eqref{U} to the initial state and obtain the following three-qubit state
\begin{eqnarray}\label{Wp}
	\begin{aligned}
		|\psi\rangle_{W'} =&\frac{1}{\sqrt{3}}(\ket{\tilde{1}\tilde{0}\tilde{0}}+\ket{\tilde{0}\tilde{1}\tilde{0}}+\ket{\tilde{0}\tilde{0}\tilde{1}})\\
		=& \frac{1}{\sqrt{24}}(3|000\rangle+|001\rangle+|010\rangle-|011\rangle+|100\rangle \\& 
		-|101\rangle-|110\rangle+|111\rangle),
	\end{aligned}
\end{eqnarray}
where the new basis is defined as $\ket{\tilde{0}}=(\ket{0}+\ket{1})/\sqrt{2}$ and $\ket{\tilde{1}}=(\ket{0}-\ket{1})/\sqrt{2}$, and the state $\ket{\psi}_{W'}$ is equivalent to the standard W state in Eq. \eqref{W} up to local unitary operations. In strategy MS-VI, the POVM operators for Alice and Bob take the form
\begin{equation}
\begin{aligned}
	A_{0|0}=\frac{\mathbb{I}+\sin(\theta)\sigma_z+\cos(\theta)\sigma_x}{2},\\ A_{0|1}=\frac{\mathbb{I}+\sin(\theta)\sigma_z-\cos(\theta)\sigma_x}{2},\\
	B_{0|0}=\frac{\mathbb{I}+\sin(\theta)\sigma_z+\cos(\theta)\sigma_x}{2},\\ 
	B_{0|1}=\frac{\mathbb{I}+\sin(\theta)\sigma_z-\cos(\theta)\sigma_x}{2},
\end{aligned}
\end{equation}
where the parameter $\theta$ ranges in the interval $[0,\frac{\pi}{2}]$. The POVM operators for the observer Charlie$_k$ can be written as
\begin{equation}
		C_{0|0}^{(k)}=\frac{\mathbb{I}+\sigma_x}{2},\quad C_{0|1}^{(k)}=\frac{\mathbb{I}+\gamma_k\sigma_z}{2},
\end{equation}
where the sharpness parameter satisfies $\gamma_k\in[0,1]$, and the index of Charlie is $k=1, \dots, n$ for any fixed $n\in \mathbb{N}$. In Appendix C, we prove that the unbounded sequential Mermin nonlocality sharing in the noiseless scenario with the standard W state can be realized using the strategy MS-VI assisted by local unitary operations.

We now study the robustness of SSQN against noise in local quantum channels. As shown in Fig. \ref{f3}, when Alice, Bob, and Charlie$_{1}$ initially share the standard W state in Eq. \eqref{W}, we consider strategy MS-V and the local quantum channel is subject to phase-flip noise. Then, under the influence of the local noisy channel $\varepsilon_1$ defined in Eq. \eqref{bit}, we obtain the expected Mermin value for the observers Alice, Bob and Charlie$_k$
\begin{equation}\label{MerminW}
\begin{aligned}	
	I^{(k)}_{\mathcal{M}}(\varepsilon_1)_{\mathrm{MS\text{-}V}}=&\frac{P_{k}}{2^{k-2}}\left(\cos^2\theta+\frac{2\sin^2\theta}{3}\right)\\ &+ \frac{\gamma_k(2p-1)^{k-1}}{2^{k-2}}\left(\frac{4\sin\theta \cos\theta }{3}\right),
\end{aligned}
\end{equation}
where $P_{k}(\theta)=\prod\limits_{j=1}^{k-1}(1+\sqrt{1-\gamma_{j}^{2}})$ and the noise parameter satisfying $p>1/2$. To ensure a violation of the Mermin inequality $I^{(k)}_{\mathcal{M}}(\varepsilon_1)_{\mathrm{MS\text{-}V}}>2$, the sharpness parameter must satisfy $\gamma_k>[2^{k-1}-P_{k}(\cos^2\theta+\frac{2\sin^2\theta}{3})]/[(2p-1)^{k-1} 4\sin\theta \cos\theta/3]$. By introducing a parameter $\epsilon>0$, we construct a new sequence $\{\gamma_k(\theta)\}$ taking the form
\begin{equation}
	\gamma_{k}(\theta)\!=\!\left\{
	\begin{array}{ll}
		\!(1\!+\!\epsilon)(\frac{\tan(\theta)}{4}), & \hbox{$k\!=\!1$;} \\
		\!(1\!+\!\epsilon)\frac{2^{k\!-\!1}\!-\!P_{k}Q}{\frac{4\sin\theta  \cos\theta}{3}(2p\!-\!1)^{k\!-\!1}} , & \hbox{$0\!\leq\!\gamma_{k-1}(\theta)\!\leq\!1$ \!;} \\
		\infty, & \hbox{otherwise,}
	\end{array}
	\right.
\end{equation}
where $Q=\cos^2\theta+(2\sin^2\theta)/3$. Along similar lines, the Mermin expectation values can be derived for the bit-flip and depolarizing channels under strategy MS-V, as well as those for the three noisy channels under strategy MS-VI. Specifically, when strategy MS-VI assisted by local unitary operations is applied to the bit-flip noisy channel $\varepsilon_2$ with $p>1/2$, the expected Mermin value $I^{(k)}_{\mathcal{M}}(\varepsilon_2)_{\mathrm{MS\text{-}VI}}$ takes the same expression as that for the phase-flip channel under strategy MS-V given in \eqref{MerminW}. After analyzing the properties of the sequences $\{\gamma_k(\theta)\}$, we present the following theorem.

%%%%%%%%%%%%%%%%%%%%%%%%%%%%%%%%%%%%%%%%%%%%%%%%%%%%%%%%%%%%%%%%%%%%%%%%%%%%%%%%%%
\textit{Theorem 3.}---Starting from an initial W state, for each $n\in \mathbb{N}$, there exists a sequence  $\{\gamma_{k}(\theta)\}_{k=1}^{n}$ and a $\theta_n\in(0,1)$ for strategy MS-V such that $I^{(k)}_{\mathcal{M}}(\varepsilon_1)_{\mathrm{MS\text{-}V}}>2$ for all $k=1,2,\cdots, n$ under phase-flip noise with the strength $p>1/2$. Unbounded sequential Mermin nonlocality sharing is achievable under bit-flip noise $\varepsilon_2$ ($p>1/2$) via strategy MS-VI assisted by the local unitary $U_L$, whereas it is fundamentally impossible under depolarizing noise $\varepsilon_3$ for both strategies.
%%%%%%%%%%%%%%%%%%%%%%%%%%%%%%%%%%%%%%%%%%%%%%%%%%%%%%%%%%%%%%%%%%%%%%%%%%%%%%%%%%

\begin{figure}
	\epsfig{figure=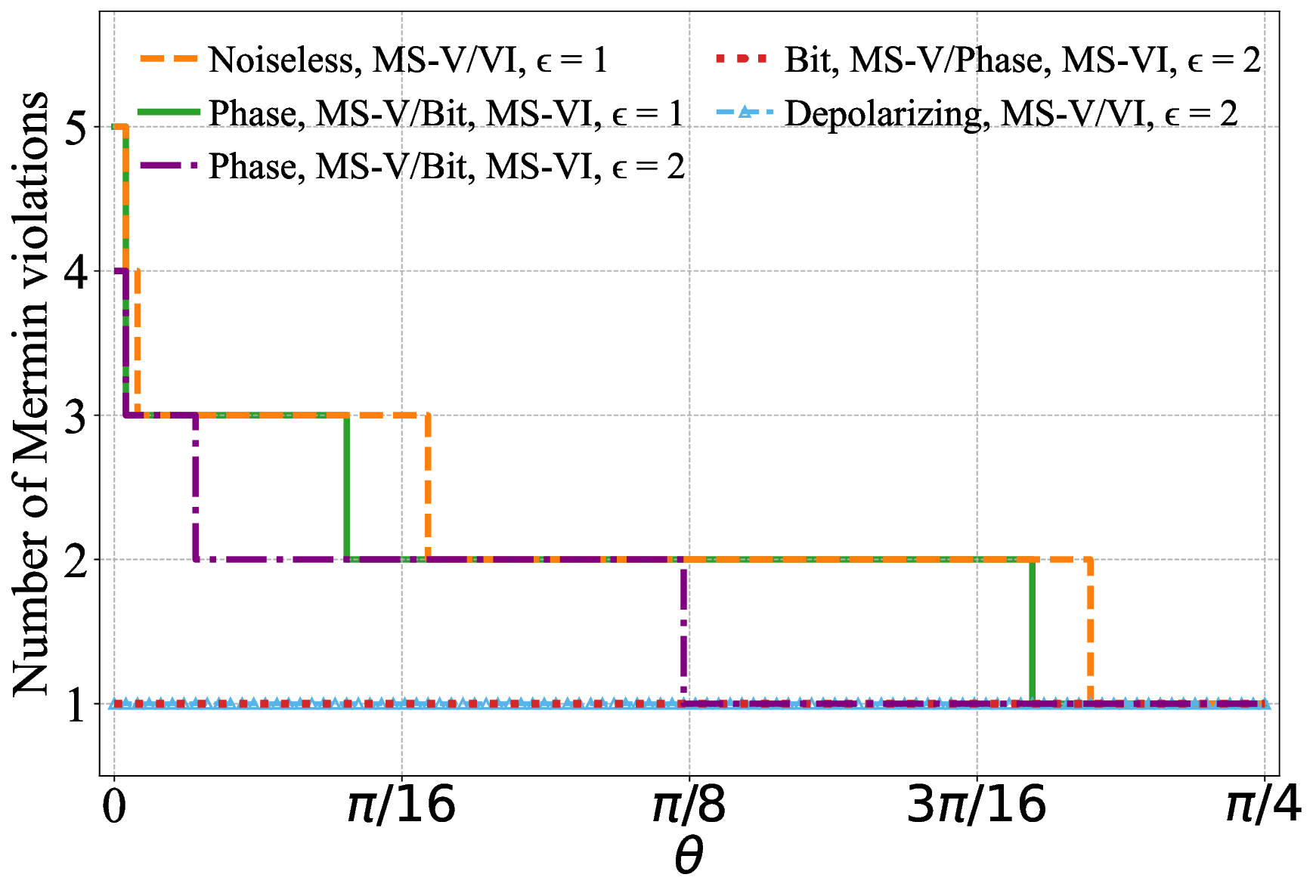,width=0.48\textwidth}
	\caption{The number of Charlies violating the Mermin inequality as a function of $\theta$ for strategies MS-V and MS-VI under noiseless and noisy quantum channels, where Alice, Bob, and Charlie$_1$ initially share a W state and the noise parameter is set to $p=0.9$ for phase-flip, bit-flip and depolarizing channels. By adaptively switching between strategies MS-V and MS-VI with appropriately chosen measurement parameters, the SSQN protocol with the W state exhibits full robustness against phase-flip and bit-flip noise, whereas it fails under depolarizing noise.}
	\label{f5}
\end{figure}

The proof of the above Theorem is provided in Appendix C, which shows that the complete noise robustness of SSQN with the W state against phase-flip and bit-flip noise can be attained by suitably switching between the two strategies. Note that the unbounded SSQN can also be achieved when the noise parameter $p<1/2$ for $\varepsilon_1$ and $\varepsilon_2$, since we can derive the same expression as that in Eq. \eqref{MerminW} via an equivalent initial state. However, when the noise parameter is $p=1/2$ for the two quantum channels, the unbounded SSQN breaks down due to the output state being a separable state. Therefore, for the given value of the noise parameter $p$ and the number $n\in \mathbb{N}$ of Charlies, the maximal number of Charlies that can sequentially share Mermin nonlocality is determined by the parameters $\theta$ and $\epsilon$, as illustrated in Fig \ref{f5} where we set $p=0.9$ and $n=5$. The orange dashed line indicates the result of SSQN in the noiseless scenario for a sequence of five Charlies, which is achieved by adjusting the parameter $\theta$ for a fixed $\epsilon=1$ under both strategies MS-V and MS-VI. The solid green line reveals that, for the noise-immune channels $\varepsilon_1$ and $\varepsilon_2$, the maximal number $n=5$ of Charlies in the SSQN protocol is realized by switching to the appropriate strategy. When $\epsilon=2$ and $\theta\geq 1.5\time 10^{-7}$, the SSQN under the two noise-immune channels can be realized for at most four Charlies (the purple dot-dashed line). Moreover, SSQN fails in all other noisy scenarios, as indicated by the red dotted line and the blue dashed line with triangle markers.

%%%%%%%%%%%%%%%%%%%%%%%%%%%%%%%%%%%%%%%%%%%%%%%%%%%%%%%%%%%%%%%%%%%%%%%%%%%%%%%%%%%%%%%%%%%%%%%%%%%%%%%%
%%%%%%%%%%%%%%%%%%%%%%%%%%%%%%%%%%%%%%%%%%%%%%%%%%%%%%%%%%%%%%%%%%%%%%%%%%%%%%%%%%%%%%%%%%%%%%%%%%%%%%%%

\section{Noise resilience of double-violation SSQNs in noisy quantum channels}

In the noiseless scenario, the SSQNs with two sequential local observers at one side have been experimentally demonstrated in photonic systems. Based on the weak measurement strategy of Ref. \cite{rs15prl}, the double-violation of the Bell-CHSH inequality with a two-photon Bell state was experimentally realized \cite{ms17qst,mjh18njp,tff20pra}. Moreover, by employing the projective-measurement strategy assisted by classical randomness \cite{as22prl,ssa24prl,wen25qinp}, the sequential sharing of bipartite Bell and tripartite Mermin nonlocality were observed using two-qubit Bell state and three-qubit GHZ state \cite{yys24njp,xue26prl}. However, these protocols do not account for channel noise. Therefore, it is worth investigating the noise resilience of the double-violation SSQN protocol in the noisy scenario. The phase-flip, bit-flip, and depolarizing channels can be effectively simulated in photonic quantum systems \cite{Xu10nc}. In this section, we present two experimental schemes to verify the noise robustness of the double-violation SSQN using a two-qubit Bell state and a three-qubit GHZ state, respectively. 

%%%%%%%%%%%%%%%%%%%%%%%%%%%%%%%%%%%%%%%%%%%%%%%%%%%%%%%%%%%%%%%%%%%%%%%%%%%%%%%%%%%%%%%%%%%%%%%%%%%%%%%%
\begin{figure}
	\epsfig{figure=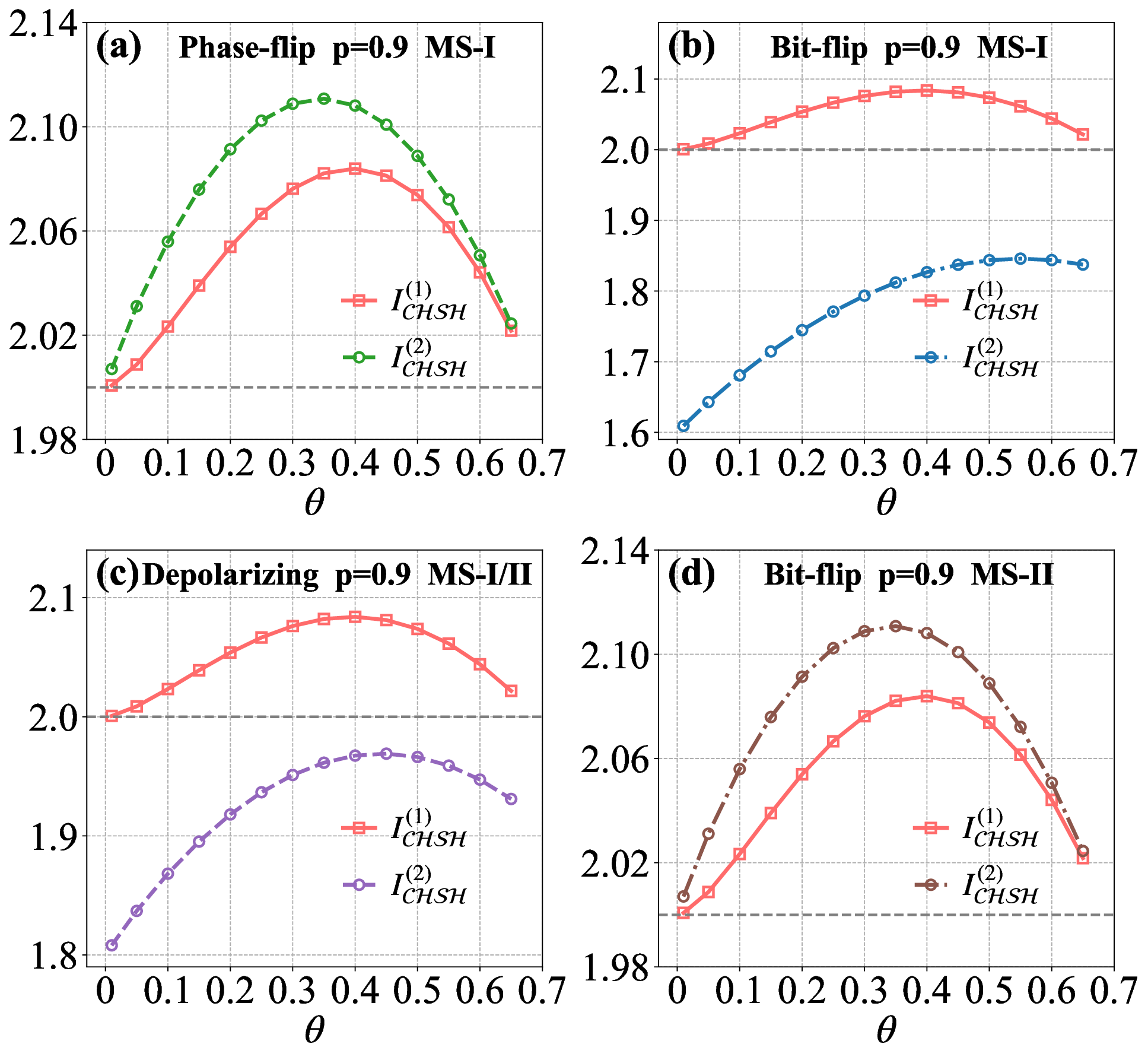,width=0.48\textwidth}
	\caption{Noise resilience of sequential Bell-CHSH nonlocality sharing with a Bell state via strategies MS-I and MS-II. The double-violation SSQN under the phase-flip and bit-flip noisy channels is achievable by adaptively switching between these strategies, but fails under depolarizing noise.	(a) The phase-flip quantum channel with strategy MS-I. (b) The bit-flip quantum channel with strategy MS-I. (c) The depolarizing quantum channel with strategies MS-I and MS-II. (d) The bit-flip quantum channel with strategy MS-II. In all panels, the noise strength is set to $p=0.9$ and the gray dashed line indicates the classical bound $I_{\mathcal{CHSH}}=2$.}
	\label{f6}
\end{figure}
%%%%%%%%%%%%%%%%%%%%%%%%%%%%%%%%%%%%%%%%%%%%%%%%%%%%%%%%%%%%%%%%%%%%%%%%%%%%%%%%%%%%%%%%%%%%%%%%%%%%%%%%

Following Theorem 1 in Sec. III, we first investigate the noise resilience of sequential Bell-CHSH nonlocality sharing with a Bell state in the noisy scenario. For the double-violation case, we set the number of Bobs to $n=2$ and fix the noise strength to $p=0.9$ for the local noisy quantum channels $\varepsilon_1$, $\varepsilon_2$ and $\varepsilon_3$ in Eqs. \eqref{phase}-\eqref{dep}. The expected CHSH values $I_{\mathcal{CHSH}}^{(k)}$ for $k=1, 2$ are functions of the measurement orientation $\theta$ and the sharpness parameter $\gamma_k$, where the sharpness of Bob$_2$ is set to $\gamma_2=1$ for the double-violation case. In Fig. \ref{f6}(a), we consider the phase-flip quantum channel $\varepsilon_1$ and plot the two Bell-CHSH correlations $I_{\mathcal{CHSH}}^{(1)}$ (the red line) and $I_{\mathcal{CHSH}}^{(2)}$ (the green line) as the functions of the parameter $\theta$, where strategy MS-I is employed and sequential Bell-CHSH nonlocality sharing under the phase-flip noise is realized. In strategy MS-I, the sharpness $\gamma_1$ is optimized by maximizing $\epsilon$ for each given value of $\theta$ based on Eqs. \eqref{bellCHSH}-\eqref{17} (the detailed analysis and the optimal parameter values are provided in Appendix D). However, under strategy MS-I with the same configuration of parameters $\theta$ and $\gamma_1$, the double-violation SSQNs are destroyed by bit-flip and depolarizing noise, as shown in Fig. \ref{f6}(b) and \ref{f6}(c). It is noted that, even when strategy MS-II is employed, the sequential Bell-CHSH nonlocality sharing is still impossible under depolarizing noise as illustrated in Fig. \ref{f6}(c), where the property $I_{\mathcal{CHSH}}^{(k)}(\varepsilon_3)_{\mathrm{MS\text{-}I}}=I_{\mathcal{CHSH}}^{(k)}(\varepsilon_3)_{\mathrm{MS\text{-}II}}$ is used. In contrast to Fig. \ref{f6}(b) with strategy MS-I, the double-violation SSQN under bit-flip noise is successfully achieved by switching to strategy MS-II as shown in Fig. \ref{f6}(d).

%%%%%%%%%%%%%%%%%%%%%%%%%%%%%%%%%%%%%%%%%%%%%%%%%%%%%%%%%%%%%%%%%%%%%%%%%%%%%%%%%%%%%%%%%%%%%%%%%%%%%%%%
\begin{figure}
	\epsfig{figure=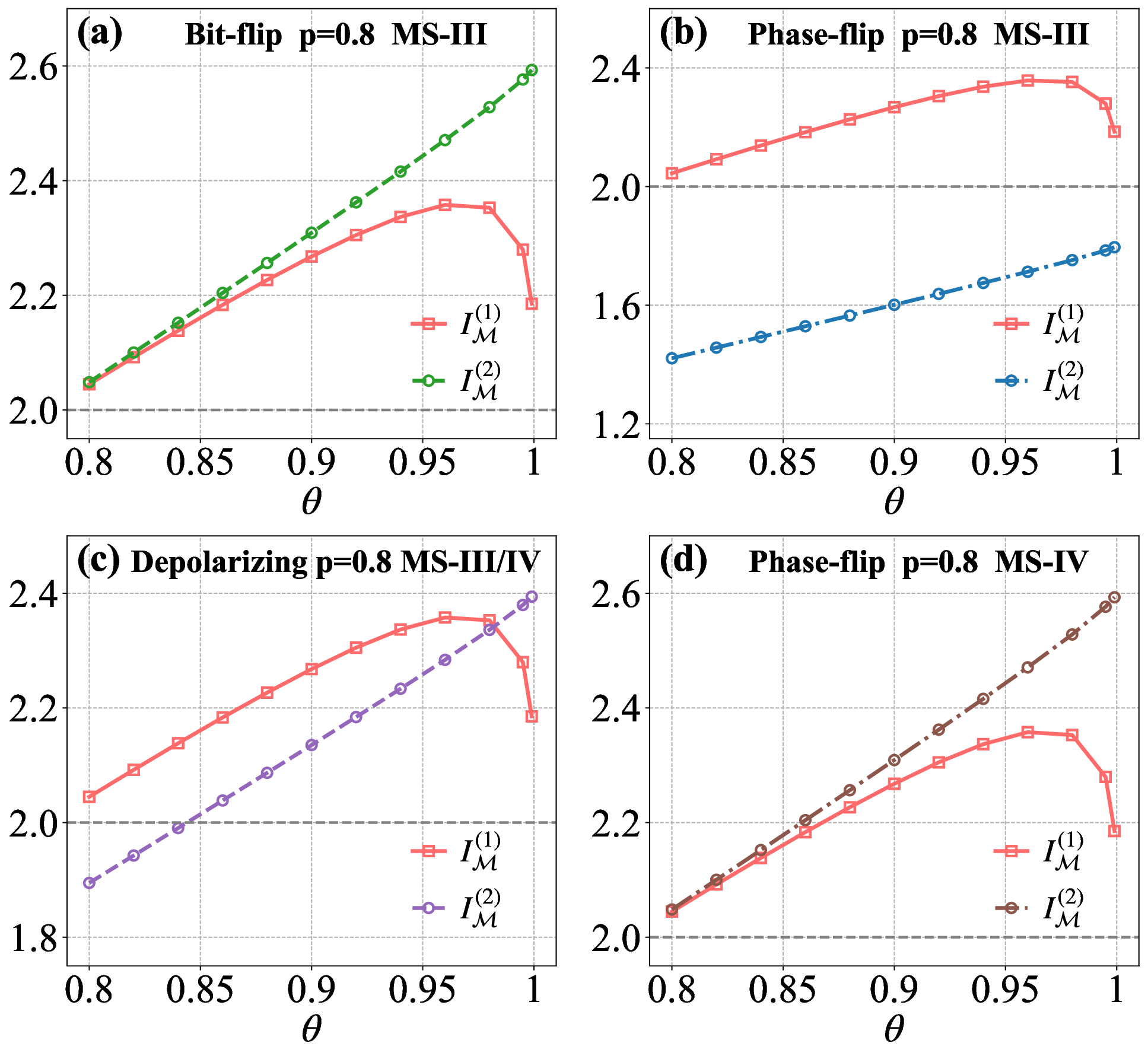,width=0.48\textwidth}
	\caption{The noise resilience of sequential tripartite Mermin nonlocality sharing schemes with a GHZ state by employing the strategies MS-III and MS-IV. (a) The bit-flip quantum channel with the strategy MS-III. (b) The phase-flip quantum channel with the strategy MS-III. (c) The depolarizing quantum channel with the strategies MS-III and MS-IV. (d) The phase-flip quantum channel with the strategy MS-IV. In each panel, the noise parameter is set to $p=0.8$ and the gray dashed line indicates the classical bound $I_{\mathcal{M}}=2$.}
	\label{f7}
\end{figure}
%%%%%%%%%%%%%%%%%%%%%%%%%%%%%%%%%%%%%%%%%%%%%%%%%%%%%%%%%%%%%%%%%%%%%%%%%%%%%%%%%%%%%%%%%%%%%%%%%%%%%%%%

For the tripartite settings, we investigate the noise resilience of sequential Mermin nonlocality sharing with a GHZ state in the noisy scenario. Following Theorem 2 in Sec. IV A, we set the number of Charlies to $n=2$ and fix the noise parameter to $p=0.8$ for all noisy channels. In this case, the expected Mermin values $I_{\mathcal{M}}^{(k)}$ for $k=1, 2$ are functions of parameters $\theta$ and $\gamma_1$, where the sharpness value of Charlie$_2$ is set to $\gamma_2=1$. The value of $\gamma_1$ is optimized by maximizing $\epsilon$ for each given $\theta$ according to Eqs. \eqref{IM b G}-\eqref{30}. In Fig. \ref{f7}(a), we consider the bit-flip noise channel and plot the Mermin values $I_{\mathcal{M}}^{(1)}$ (the red line) and $I_{\mathcal{M}}^{(2)}$ (the green line) as functions of the parameter $\theta$, where strategy MS-III is adopted and the double-violation SSQN is successfully realized. For strategy MS-III, the optimal values of $\theta$, $\epsilon$ and $\gamma_1$ are listed in Appendix D. However, under phase-flip noise, the SSQN under strategy MS-III fails when the same parameter configuration is used, as shown in Fig. \ref{f7}(b). Similarly, under depolarizing noise, the double-violation SSQN can only be realized for sufficiently large $\theta$, irrespective of the measurement strategy, as shown in Fig. \ref{f7}(c). Moreover, in contrast to that in Fig. \ref{f7}(b), the double-violation SSQN can be successfully implemented by switching to strategy MS-IV assisted by local unitary operations in Eqs. \eqref{U}-\eqref{27}, as illustrated in Fig. \ref{f7}(d).

\section{Discussion and conclusion}

Quantum nonlocality serves as a fundamental resource for device-independent quantum information processing. Under the influence of channel noise, the sequential sharing protocols of bipartite Bell-CHSH and tripartite Mermin nonlocality provide a practical and effective method for sharing limited quantum resources among multiple observers. It is well established that quantum resources in multipartite systems, such as entanglement and quantum nonlocality, cannot be freely shared \cite{osborne06prl,hiro07prl,bai09pra,bai13pra,bai14prl,scarani01prl,cheng17prl,li25prb}. Therefore, a particularly interesting question is whether the total achievable Bell violations in both noiseless and noisy SSQN scenarios is subject to a monogamy constraint. Moreover, beyond the SSQN, sequential sharing protocols have been extensively explored for a wide range of quantum information resources, including preparation contextuality \cite{aa19pra,hnr21qtm,aa23pra}, quantum entanglement \cite{cmu22pra,mh23pra,cmu25pra}, quantum steering \cite{sds18pra,sh19pra,cyh20optica,dc21pra,szc24pra}, nonlocal advantage of quantum coherence \cite{sa18pra}, and network nonlocality \cite{sa22pra,tns23fp,hfh24pra,hfh25pra}. In practical environments, these protocols are inevitably affected by noise channels. Consequently, investigating the impact of such noise on the shareability of different quantum resources is a problem of direct practical relevance.

In summary, we have studied the sequential sharing of bipartite Bell-CHSH and tripartite Mermin nonlocality in the presence of noisy quantum channels. We prove that the strategy MS-I \cite{pjb20prl}, originally designed for the noiseless scenario, enables unbounded SSQN evaluated by the CHSH inequality under phase-flip noise, but fails under bit-flip and depolarizing noise. Notably, our proposed strategy MS-II restores this unbounded shareability in the presence of bit-flip noise. For tripartite Mermin nonlocality, we demonstrate that unbounded SSQN via strategy MS-III \cite{yx23pra} with a GHZ state resists bit-flip noise, whereas the W-state protocol employing MS-V \cite{bz24pra} endures phase-flip noise. In these tripartite scenarios, we further introduce local-unitary-assisted strategies MS-IV and MS-VI, which allow the noise immunity to be actively switched between different noisy channels. In addition, we propose double-violation SSQN schemes for both bipartite and tripartite settings, demonstrating the noise resilience of these protocols under various measurement strategies.  Our findings provide key insight into the behavior of SSQN under noisy quantum channels and reveal the relationship between noise robustness and adaptive measurement strategies.

\section*{Acknowledgments}
This work was supported by NSF-China (Grants Nos. 11575051, 12404330, 12404405, and 12105074), Hebei NSF (Grant No. A2021205020),  Hebei 333 Talent Project (No. B20231005), Science and Technology Project of Hebei Education Department (Grant No. BJ2025036), and the fund of Hebei Normal University (Grant No. L2026J02). W.-L.M. also acknowledges support from the National Natural Science Foundation of China (Grants No. 12574082 and No. E31Q02BG), the Chinese Academy of Sciences (Grants No. E0SEBB11 and No. E27RBB11), the Innovation Program for Quantum Science and Technology (Grant No. 2021ZD0302300), and the Chinese Academy of Sciences Project for Young Scientists in Basic Research (YSBR-090).

\section*{Data availability}

The data underlying the findings reported in this article are not publicly available. The data available from the authors upon reasonable request.

\onecolumngrid

\appendix

%%%%%%%%%%%%%%%%%%%%%%%%%%%%%%%%%%%%%%%%%%%%%%%%%%%%%%%%%
\section{The proof of Theorem 1}	
We first consider the SSQN protocol for the Bell state by employing the strategy MS-I in Eqs.  \eqref{bellA}-\eqref{bellB}. Assume that the local channel is subject to phase-flip noise, \textit{i.e.}, the noisy channel $\varepsilon_1$ given in Eq. \eqref{phase}. Let $\rho_{Bell}^{(k-1)}(\varepsilon_1)$ denote the quantum state shared by Bob$_{k-1}$ and Alice, prior to Bob$_{k-1}$'s measurement. Then, after Bob$_{k-1}$ performs his measurement, the post-measurement state according to L\"{u}ders rule can be written as
\begin{equation}\label{A1}
	\rho_{Bell}^{(k-1)'} = \frac{2+\sqrt{1-\gamma_{k-1}^{2}}}{4}\rho_{Bell}^{(k-1)}(\varepsilon_1)+\frac{1}{4}(\mathbb{I}\otimes\sigma_{z})\rho_{Bell}^{(k-1)}(\varepsilon_1)(\mathbb{I}\otimes\sigma_{z})
	+\frac{1-\sqrt{1-\gamma_{k-1}^{2}}}{4}(\mathbb{I}\otimes\sigma_{x})\rho_{Bell}^{(k-1)}(\varepsilon_1)(\mathbb{I}\otimes\sigma_{x}).
\end{equation}
The post-measurement qubit is further relayed to Bob$_{k}$ through the noisy quantum channel $\varepsilon_1$, and the resulting quantum state shared by Bob$_k$ and Alice takes the form
\begin{equation}\label{A2}
	\begin{aligned}
		\rho_{Bell}^{(k)}(\varepsilon_1) &=p\rho_{Bell}^{(k)'}+(1-p)\left(\mathbb{I}\otimes\sigma_z\rho_{Bell}^{(k-1)'}\mathbb{I}\otimes\sigma_z\right)\\
		&=p\frac{2\!\!+\!\!\sqrt{1\!-\!\gamma_{k-1}^{2}}}{4}\rho_{Bell}^{(k-1)}(\varepsilon_1)+\frac{p}{4}(\mathbb{I}\otimes\sigma_{z})\rho_{Bell}^{(k-1)}(\varepsilon_1)(\mathbb{I}\otimes\sigma_{z})
		+p\frac{1\!\!-\!\!\sqrt{1\!-\!\gamma_{k-1}^{2}}}{4}(\mathbb{I}\otimes\sigma_{x})\rho_{Bell}^{(k-1)}(\varepsilon_1)(\mathbb{I}\otimes\sigma_{x})\\
		&\hspace{1em}+(1\!-\!p)\frac{2\!+\!\sqrt{1\!-\!\gamma_{k-1}^{2}}}{4}(\mathbb{I}\otimes\sigma_{z})\rho_{Bell}^{(k-1)}(\varepsilon_1)(\mathbb{I}\otimes\sigma_{z})
		+\frac{1\!-\!p}{4}(\mathbb{I}\otimes\sigma_{z})(\mathbb{I}\otimes\sigma_{z})\rho_{Bell}^{(k-1)}(\varepsilon_1)
		(\mathbb{I}\otimes\sigma_{z})(\mathbb{I}\otimes\sigma_{z})\\[3mm]
		&\hspace{1em}+(1\!-\!p)\frac{1\!-\!\sqrt{1\!-\!\gamma_{k-1}^{2}}}{4}(\mathbb{I}\otimes\sigma_{z})(\mathbb{I}\otimes\sigma_{x})\rho_{Bell}^{(k-1)}(\varepsilon_1)
		(\mathbb{I}\otimes\sigma_{x})(\mathbb{I}\otimes\sigma_{z}).
	\end{aligned}
\end{equation}
According to Eq. (\ref{eq:CHSH}) of main text, we derive the expected CHSH value for the quantum state $\rho_{Bell}^{(k)}(\varepsilon_1)$
\begin{equation}\label{Ik p bell}
	\begin{aligned}
		I^{(k)}_{\mathcal{CHSH}}(\varepsilon_1)_{\mathrm{MS\text{-}I}} & = \mathrm{Tr}\left[\rho_{Bell}^{(k)}(\varepsilon_1)(X_{0}Y^{(k)}_{0}+X_{0}Y^{(k)}_{1}+X_{1}Y^{(k)}_{0}-X_{1}Y^{(k)}_{1})\right]\\[4mm]
		& =  2^{2-k}\left[\gamma_k\sin(\theta)(2p-1)^{k-1}+\cos(\theta)\prod_{j=1}^{k-1}(1+\sqrt{1-\gamma_j^2})\right].
	\end{aligned}
\end{equation}
In the calculation of the expectation values in Eq. \eqref{Ik p bell}, we use the following expressions
\begin{eqnarray*}\label{rl}
\mathrm{Tr}\left[\rho_{Bell}^{(k)}(\varepsilon_1)X_{0}Y^{(k)}_{0}\right]&=&\mathrm{Tr}\left[\rho_{Bell}^{(k)}(\varepsilon_1)\left(\cos(\theta)\sigma_z+\sin(\theta)\sigma_x\right)\otimes\sigma_z\right]=2^{1-k}\left[\cos(\theta)\prod_{j=1}^{k-1}\left(1+\sqrt{1-\gamma_j^2}\right)\right],\\
	\mathrm{Tr}\left[\rho_{Bell}^{(k)}(\varepsilon_1)X_{1}Y^{(k)}_{0}\right]&=&\mathrm{Tr}\left[\rho_{Bell}^{(k)}(\varepsilon_1)\left(\cos(\theta)\sigma_z-\sin(\theta)\sigma_x\right)\otimes\sigma_z\right]=2^{1-k}\left[\cos(\theta)\prod_{j=1}^{k-1}\left(1+\sqrt{1-\gamma_j^2}\right)\right],\\
\mathrm{Tr}\left[\rho_{Bell}^{(k)}(\varepsilon_1)X_{0}Y^{(k)}_{1}\right]&=&\mathrm{Tr}\left[\rho_{Bell}^{(k)}(\varepsilon_1)\left(\cos(\theta)\sigma_z+\sin(\theta)\sigma_x\right)\otimes(\gamma_k\sigma_x)\right]=2^{1-k}\left[\gamma_k\sin(\theta)(2p-1)^{k-1}\right],\\
	-\mathrm{Tr}\left[\rho_{Bell}^{(k)}(\varepsilon_1)X_{1}Y^{(k)}_{1}\right]&=&\mathrm{Tr}\left[\rho_{Bell}^{(k)}(\varepsilon_1)\left(\cos(\theta)\sigma_z-\sin(\theta)\sigma_x\right)\otimes(\gamma_k\sigma_x\right]=2^{1-k}\left[\gamma_k\sin(\theta)(2p-1)^{k-1}\right].
\end{eqnarray*}
In particular, we have the expected CHSH value for $k=1$, which can be written as $I^{(1)}_{\mathcal{CHSH}}(\varepsilon_1)_{\mathrm{MS\text{-}I}}=2(\gamma_{1}\sin(\theta)+\cos(\theta))$.
To ensure that $I^{(k)}_{\mathcal{CHSH}}(\varepsilon_1)_{\mathrm{MS\text{-}I}}>2$, we define a new sequence $\{\gamma_{k}(\theta)\}$ with $\epsilon>0$,
\begin{equation}\label{gamma p bell}
	\gamma_{k}(\theta)=\left\{
	\begin{array}{ll}
		(1+\epsilon)\frac{2^{k-1}-\cos(\theta)\prod\limits_{j=1}^{k-1}(1+\sqrt{1-\gamma_{j}^{2}})}{\sin(\theta)(2p-1)^{k-1}} , & \hbox{$0\leq\gamma_{k-1}(\theta)\leq1$ ,} \\[2mm]
		\infty, & \hbox{otherwise.}
	\end{array}
	\right.
\end{equation}
Without loss of generality, the noise parameter is set to $p>1/2$ in the phase-flip quantum channel. Given the constraint $0<\gamma_{k-1}(\theta)\leq1$, we can derive the relation
\begin{equation}\label{ratio}
	\frac{\gamma_{k}(\theta)}{\gamma_{k-1}(\theta)}>2.
\end{equation}

Next, we utilize the inequalities $\sqrt{1-x^2}\geq1-x^2$ for $x\in[0,1]$, $\cos(\theta)\geq1-\theta^2/2$ and $\sin(\theta)\geq\theta/2$ for $\theta\in(0,\pi/4]$. After applying these inequalities to $\gamma_{k}(\theta)$ in Eq. \eqref{gamma p bell}, we find that, when  $\gamma_{k}(\theta)$ is finite, the following relation holds
\begin{equation}\label{rrk}
	\gamma_{k}(\theta) \leq(1+\epsilon)\frac{2^{k-1}-(1-\theta^2/2)\prod\limits_{j=1}^{k-1}(2-\gamma_{j}^{2}(\theta))}
	{\theta/2(2p-1)^{k-1}}
	=(1+\epsilon)2^k\frac{1-(1-\theta^2/2)\prod\limits_{j=1}^{k-1}(1-\gamma_{j}^{2}(\theta)/2)}
	{\theta(2p-1)^{k-1}}.
\end{equation}
Then, we define a new sequence based on the upper bound in Eq. \eqref{rrk}, which has the form
\begin{equation}\label{qk}
	q_{k}(\theta)=\left\{
	\begin{array}{ll}
		(1+\epsilon)2^k\frac{1-(1-\theta^2/2)\prod\limits_{j=1}^{k-1}\left(1-q_{j}^{2}(\theta)/2\right)}
		{\theta(2p-1)^{k-1}} , & \hbox{$0\leq q_{k-1}(\theta)\leq1$ ,} \\[2mm]
		\infty, & \hbox{otherwise,}
	\end{array}
	\right.
\end{equation}
where the quantity $q_{1}(\theta)=(1+\epsilon)\theta$. Due to $q_k(\theta) \geq \gamma_k(\theta)$, the finiteness of $\gamma_k(\theta)$ is guaranteed, whenever $q_k(\theta)$ is finite. Moreover, the condition $\frac{\gamma_{k}(\theta)}{\gamma_{k-1}(\theta)}>2$ in Eq. \eqref{ratio} implies that $q_{k}$ is monotonically increasing along with the index $k$.

We now prove by induction that for each $k \in \mathbb{N}$, there exists $\theta_{k} \in (0,\pi/4]$ such that $0 < q_{k}(\theta) < 1$ for all $\theta \in (0,\theta_{k})$ and $\lim_{\theta \to 0^{+}}q_{k}(\theta) = 0$.

For $k = 1$, we observe that $q_{1}(\theta)$ is a polynomial in $\theta$ with odd degree, and satisfies $0 < q_{1}(\theta) < 1$ when $\theta \in (0,\frac{1}{1+\varepsilon})$. Thus, $q_{2}(\theta)$ is an odd polynomial with no constant term on the interval $(0,\frac{1}{1+\varepsilon})$, and then we can obtain $\lim_{\theta \to 0^{+}}q_{2}(\theta) = 0$. Due to the property $q_{2}(\theta)>q_{1}(\theta)>0$, $q_{2}(\theta)$ is a continuous positive function on the interval, and we have  $0 < q_{2}(\theta) < 1$ for $\theta \in (0,\theta_{2})$  with $\theta_2 \in (0,\frac{1}{1+\varepsilon})$. Assume that there exists $\theta_{k-1} > 0$ such that, for $j = 1,\ldots,k-1$,   $q_{j}(\theta)$s are odd polynomials in $\theta$ and satisfy $0 < q_{j}(\theta) < 1$ when $\theta \in(0,\theta_{k-1})$. Then, on this interval, the numerator of $q_{k}(\theta)$ is an even polynomial in $\theta$ without a constant term. After canceling the $\theta$ in the denominator, $q_{k}(\theta)$ becomes an odd polynomial, indicating that $\lim_{\theta \to 0^{+}}q_{k}(\theta) = 0$. Since $q_{k}(\theta)>0$ and $q_{k}(\theta)$ being a continuous positive function,  there exists $\theta_{k} \in (0,\theta_{k-1})$ such that $0 < q_{k}(\theta) < 1$ for all $\theta \in (0,\theta_{k})$.

By induction, for any $n \in \mathbb{N}$, we can find a $\theta_{n} \in (0,\pi/4]$ such that $0 < q_{1}(\theta) < \cdots < q_{n}(\theta) < 1$ for all $\theta \in (0,\theta_{n})$. Since $0 < \gamma_{k}(\theta) \leq q_{k}(\theta)$, it follows that $0 < \gamma_{1}(\theta) < \cdots < \gamma_{n}(\theta) < 1$ on $(0,\theta_{n})$ and $\lim_{\theta \to 0^{+}}\gamma_{n}(\theta) = 0$. Therefore, in the presence of phase-flip noise, a single Alice can share Bell-CHSH nonlocality with arbitrarily many independent Bobs via the measurement strategy MS-I.

Next, we consider the case of strategy MS-II shown in Eq. \eqref{bellAB} under the bit-flip channel $\varepsilon_2$. Let $\rho_{\text{Bell}}^{(k-1)}(\varepsilon_2)$ be the state shared by Alice and Bob$_{k-1}$ before his measurement, after which the post-measurement state reads
\begin{equation}
	\rho_{Bell}^{(k-1)''} = \frac{2+\sqrt{1-\gamma_{k-1}^{2}}}{4}\rho_{Bell}^{(k-1)}(\varepsilon_2)+\frac{1}{4}(\mathbb{I}\otimes\sigma_{x})\rho_{Bell}^{(k-1)}(\varepsilon_2)(\mathbb{I}\otimes\sigma_{x})
	+\frac{1-\sqrt{1-\gamma_{k-1}^{2}}}{4}(\mathbb{I}\otimes\sigma_{z})\rho_{Bell}^{(k-1)}(\varepsilon_2)(\mathbb{I}\otimes\sigma_{z}).
\end{equation}
After the qubit is transmitted to Bob$_k$ via the noisy channel $\varepsilon_2$, the output state shared by Bob$_k$ and Alice can be written as
\begin{equation}
	\begin{aligned}
		\rho_{Bell}^{(k)}(\varepsilon_2) &=p\rho_{Bell}^{(k)''}+(1-p)\left(\mathbb{I}\otimes\sigma_x\rho_{Bell}^{(k-1)''}\mathbb{I}\otimes\sigma_x\right)\\
		&=p\frac{2\!+\!\!\sqrt{1\!-\!\gamma_{k-1}^{2}}}{4}\rho_{Bell}^{(k-1)}(\varepsilon_2)\!+\!\frac{p}{4}(\mathbb{I}\otimes\sigma_{x})\rho_{Bell}^{(k-1)}(\varepsilon_2)(\mathbb{I}\otimes\sigma_{x})
		\!+\!p\frac{1\!-\!\sqrt{1\!-\!\gamma_{k-1}^{2}}}{4}(\mathbb{I}\otimes\sigma_{z})\rho_{Bell}^{(k-1)}(\varepsilon_2)(\mathbb{I}\otimes\sigma_{z})\\
		&\hspace{1em}+(1\!-\!p)\frac{2\!+\!\sqrt{1\!-\!\gamma_{k-1}^{2}}}{4}(\mathbb{I}\otimes\sigma_{x})\rho_{Bell}^{(k-1)}(\varepsilon_2)(\mathbb{I}\otimes\sigma_{x})
		+\frac{1\!-\!p}{4}(\mathbb{I}\otimes\sigma_{x})(\mathbb{I}\otimes\sigma_{x})\rho_{Bell}^{(k-1)}(\varepsilon_2)
		(\mathbb{I}\otimes\sigma_{x})(\mathbb{I}\otimes\sigma_{x})\\
		&\hspace{1em}+(1\!-\!p)\frac{1\!-\!\sqrt{1\!-\!\gamma_{k-1}^{2}}}{4}(\mathbb{I}\otimes\sigma_{x})(\mathbb{I}\otimes\sigma_{z})\rho_{Bell}^{(k-1)}(\varepsilon_2)
		(\mathbb{I}\otimes\sigma_{z})(\mathbb{I}\otimes\sigma_{x}).
	\end{aligned}
\end{equation}
For the above shared state, the expected CHSH value under strategy MS-II is
\begin{equation}\label{Ik p bell x}
	\begin{aligned}
	I^{(k)}_{\mathcal{CHSH}}(\varepsilon_2)_{\mathrm{MS\text{-}II}} & = \mathrm{Tr}\left[\rho_{Bell}^{(k)}(\varepsilon_2)(X_{0}Y^{(k)}_{0}+X_{0}Y^{(k)}_{1}+X_{1}Y^{(k)}_{0}-X_{1}Y^{(k)}_{1})\right]\\[4mm]
		& =  2^{2-k}\left[\gamma_k\sin(\theta)(2p-1)^{k-1}+\cos(\theta)\prod_{j=1}^{k-1}\left(1+\sqrt{1-\gamma_j^2}\right)\right],
	\end{aligned}
\end{equation}
which is identical to that for the phase-flip channel under strategy MS-I as given in Eq. \eqref{Ik p bell}, indicating that the roles of the bit-flip and phase-flip channels are swapped under the two measurement strategies. Therefore, in the presence of bit-flip quantum channel $\varepsilon_2$, arbitrarily many independent Bobs can share the Bell-CHSH nonlocality of the Bell state with a single Alice under strategy MS-II.

For the remaining four combinations of measurement strategies (MS-I and MS-II) and noisy channels ($\varepsilon_i$ with $i=1,2,3$), the corresponding expected CHSH values can be derived in a similar way by using the method given in Eqs. \eqref{A1}-\eqref{Ik p bell} and we can obtain
\begin{eqnarray}\label{A10&11}
&&I^{(k)}_{\mathcal{CHSH}}(\varepsilon_2)_{\mathrm{MS\text{-}I}}=I^{(k)}_{\mathcal{CHSH}}(\varepsilon_1)_{\mathrm{MS\text{-}II}}=2^{2-k}\left[\gamma_k\sin(\theta)+\cos(\theta)\prod_{j=1}^{k-1}(2p-1)\left(1+\sqrt{1-\gamma_j^{2}}\right)\right],\label{a11}\\
&&I^{(k)}_{\mathcal{CHSH}}(\varepsilon_3)_{\mathrm{MS\text{-}I}}=I^{(k)}_{\mathcal{CHSH}}(\varepsilon_3)_{\mathrm{MS\text{-}II}}=2^{2-k}p^{k-1}\left[\gamma_k\sin(\theta)+\cos(\theta)\prod_{j=1}^{k-1}\left(1+\sqrt{1-\gamma_j^{2}}\right)\right],\label{a12}
\end{eqnarray}
where the result for the bit-flip channel under strategy MS-I is identical to that for phase-flip channel under strategy MS-II, and the expectation values of CHSH operator for the depolarizing channel have the same expression under both strategies.

Following the method given in Eqs. \eqref{gamma p bell} to \eqref{qk}, we can define the corresponding sequences $\{q_{k}(\theta)\}$ for Eqs. \eqref{a11} and \eqref{a12}. The noise parameter for the phase-flip and bit-flip channels is set to $p>1/2$ in Eq. \eqref{a11}. However, for $k\geq2$, the corresponding sequences $q_{k}(\theta)$ are no longer odd polynomials, and thus $\lim_{\theta\rightarrow0^+}q_{2}(\theta)=0$ does not hold, which renders the sharpness parameter $\gamma_k$ non-physical for certain value of $k$. Consequently, the unbounded sequential sharing of the bipartite Bell-CHSH nonlocality is not achievable in the above noise‑affected scenarios.

Therefore, in the presence of the phase-flip and bit-flip channels, the unbounded sequential sharig of Bell-CHSH nonlocality of the Bell state can be realized by selectively adopting the measurement strategies MS-I and MS-II. But the unbounded SSQN under the depolarizing channel is unattainable for either measurement strategy. The proof of Theorem 1 is completed.

\hfill$\blacksquare$

Additionally, we analyze the SSQN in the noiseless scenario under our proposed measurement strategy MS-II. By setting $p=1$, Eq. \eqref{Ik p bell x} reduces to the noiseless CHSH expectation value that is the same to that given in Ref. \cite{pjb20prl}, which indicates that unbounded SSQN can be achieved by using the strategy MS‑II in the noiseless quantum channel. 

%%%%%%%%%%%%%%%%%%%%%%%%%%%%%%%%%%%%%%%%%%%
\section{The proof of Theorem 2}
In the Mermin scenario with the initial GHZ state, following a method analogous to that in Eqs. \eqref{A1}-\eqref{A2}, the quantum state shared by Alice, Bob and Charlie$_k$ can be derived. We first analyze the case for the bit-flip channel $\varepsilon_2$ under strategy MS-III.  The expected Mermin value can be calculated according to Eq. \eqref{eq:Mermin} in the main text, which is given by
\begin{equation}\label{B1}
	\begin{aligned}
		I_{\mathcal{M}}^{(k)}(\varepsilon_2)_{\mathrm{MS\text{-}III}} & =  2\theta\left[\gamma_k(p-\frac{1}{2})^{k-1}+2^{1-k}\prod_{j=1}^{k-1}\left(1+\sqrt{1-\gamma_j^{2}}\right)\right],
	\end{aligned}
\end{equation}
where $\theta$, $\gamma_k$, and $\gamma_j$ are sharpness parameters, and the noise parameter is set to $p>1/2$. To ensure the expected Mermin value $I^{(k)}_{\mathcal{M}}(\varepsilon_2)_{\mathrm{MS\text{-}III}}>2$, we can define a sequence $\{\gamma_{k}(\theta)\}$ with $\epsilon>0$, which have the form
\begin{equation}\label{B2}
	\gamma_{k}(\theta)=\left\{
	\begin{array}{ll}
		(1+\epsilon)(\frac{1}{\theta}-1), & \hbox{$k=1$,} \\[2mm]
		(1+\epsilon)(p-\frac{1}{2})^{1-k}(\frac{1}{\theta}-2^{1-k}P_{k}), & \hbox{$0\leq\gamma_{k-1}(\theta)\leq1$,} \\[2mm]
		\infty, & \hbox{otherwise,}
	\end{array}
	\right.
\end{equation}
where $P_{k}=\prod\limits_{j=1}^{k-1}(1+\sqrt{1-\gamma_{j}^{2}})$. According to Eq. \eqref{B2}, we can obtain $\gamma_1(\theta)=(1+\epsilon)(\frac{1}{\theta}-1)$ and $\lim\limits_{\theta\rightarrow 1^{-}}\gamma_{1}(\theta)=0$. Since $0<\gamma_{k-1}(\theta)\leq1$, we can derive the relation $\frac{\gamma_{k}(\theta)}{\gamma_{k-1}(\theta)}>2$. By induction, we assume that there exists a $\theta_{k-1}$ such that on the interval $\theta\in(\theta_{k-1},1)$, $\gamma_{i}(\theta)\in(0,1)$ and $\lim\limits_{\theta\rightarrow 1^{-}}\gamma_{i}(\theta)=0$ are true for $i=1,2,\cdots,k-1$,  which implies $\lim\limits_{\theta\rightarrow 1^{-}} P_{k-1}=2^{k-2}$ and $\lim\limits_{\theta\rightarrow 1^{-}} P_k=2^{k-1}$. According to the definition of $\gamma_{k}(\theta)$, we obtain
\begin{equation}\label{lim}
\lim\limits_{\theta\rightarrow 1^{-}}\gamma_{k}(\theta)=\lim\limits_{\theta\rightarrow 1^{-}} (1+\epsilon)\left(p-\frac{1}{2}\right)^{1-k} \left(\frac{1}{\theta}-2^{1-k}P_{k}\right) =(1+\epsilon)\left(p-\frac{1}{2}\right)^{1-k}(1-1)=0.
\end{equation}
Therefore, for any $n \in \mathbb{N}$, we can find a $\theta_{n}\in(0,1)$ such that $0<\gamma_{1}(\theta)<\gamma_{2}(\theta)<\cdots<\gamma_{n}(\theta)<1$ for all $\theta\in(\theta_{n},1).$ That is to say, in the presence of bit-flip noise, the strategy MS-III enables a single Alice and a single Bob to share Mermin nonlocality of the GHZ state with arbitrarily many independent Charlies.

Next, we consider the phase-flip channel $\varepsilon_1$ under strategy MS-IV. In this case, the initial  GHZ state undergoes a local unitary transformation $U_L$, and the output state is given by
\begin{equation}\label{ghzp}
	\begin{aligned}
		|\psi\rangle_{GHZ'}&=U_A\otimes U_B\otimes U_C|\psi\rangle_{GHZ} \\
		&=\frac{1}{2} (\vert 000\rangle + \vert 011\rangle + \vert 101\rangle + \vert 110\rangle),
	\end{aligned}
\end{equation}
where $U_i$ ($i=A,B,C$) is a Hadamard gate $H = \frac{1}{\sqrt{2}} \begin{pmatrix} 1 & 1 \\ 1 & -1 \end{pmatrix}$. The resulting state $\ket{\psi}_{GHZ'}$ is known as the tetrahedral state \cite{pa06pra,ab19pra}. Because the Hadamard gate is self-inverse ($H^{-1} = H$), the same operation $U_L$ can transforms the tetrahedral state back to the standard GHZ state. After some derivation, the expected Mermin value on $|\psi\rangle_{GHZ'}$ by adopting strategy MS-IV is
\begin{equation}\label{B4}
	\begin{aligned}
		I^{(k)}_{\mathcal{M}}(\varepsilon_1)_{\mathrm{MS\text{-}IV}}  =  2\theta\left[\gamma_k(p-\frac{1}{2})^{k-1}+2^{1-k}\prod_{j=1}^{k-1}\left(1+\sqrt{1-\gamma_j^{2}}\right)\right],
	\end{aligned}
\end{equation}
which matches the result given in Eq. \eqref{B1} for the bit-flip channel under strategy MS-III. After a similar analysis, we can obtain the conclusion that when the local channel is subject to phase-flip noise, for each $n\in \mathbb{N}$, one can construct a sequence $\{\gamma_{k}(\theta)\}_{k=1}^{n}$ under the strategy MS-IV satisfying $I^{(k)}_{\mathcal{M}}(\varepsilon_1)_{\mathrm{MS\text{-}IV}}>2$  for $k=1,2,\cdots, n$. 

For the remaining combinations of strategies (MS-III and MS-IV) and noisy channels ($\varepsilon_i$ with $i=1, 2, 3$), the corresponding expected Mermin values for Alice, Bob, and Charlie$_{k}$ can be derived in a similar way, which can be written as 
\begin{eqnarray}
	&&I_{\mathcal{M}}^{(k)}(\varepsilon_1)_{\mathrm{MS\text{-}III}}=I_{\mathcal{M}}^{(k)}(\varepsilon_2)_{\mathrm{MS\text{-}IV}}=2\theta\left[\gamma_k\left(p-\frac{1}{2}\right)^{k-1}+\left(p-\frac{1}{2}\right)^{k-1}\prod_{j=1}^{k-1}\left(1+\sqrt{1-\gamma_j^{2}}\right)\right],\label{B6}\\
    &&I_{\mathcal{M}}^{(k)}(\varepsilon_3)_{\mathrm{MS\text{-}III}}=I_{\mathcal{M}}^{(k)}(\varepsilon_3)_{\mathrm{MS\text{-}IV}}=2\theta\left[\gamma_k\left(\frac{p}{2}\right)^{k-1}+\left(\frac{p}{2}\right)^{k-1}\prod_{j=1}^{k-1}\left(1+\sqrt{1-\gamma_j^{2}}\right)\right], \label{B7}
\end{eqnarray}
where $\theta$, $\gamma_k$, and $\gamma_j$ are sharpness parameters, and the noise parameter is set to $p>1/2$ in Eq. \eqref{B6}. By utilizing a similar analysis method like that for Eq. \eqref{B2}, we can define the corresponding sequences $\{\gamma_{k}(\theta)\}$. However, it is found that, under the induction hypothesis  $\lim\limits_{\theta\rightarrow1^-}\gamma_{k-1}(\theta)=0$, the sharpness parameter $\gamma_{k}(\theta)$ does not vanish in the limit $\theta\rightarrow1^-$, which renders the sharpness parameter $\gamma_k$ non-physical for certain value of the index $k$. Consequently, the unbounded property for sequential sharing of the tripartite Mermin nonlocality breaks down in the above noise-affected scenarios.

Therefore, starting from an initial standard GHZ state, for the phase-flip and bit-flip quantum channels, the unbounded sequential sharing of tripartite Mermin nonlocality can be achieved by selectively adopting the strategies MS-III and MS-IV. But the unbounded SSQN is destroyed by local depolarizing noise under either strategy. Thus, the proof of Theorem 2 is completed.

\hfill$\blacksquare$

In addition, we analyze the SSQN in the noiseless scenario by employing our proposed measurement strategy MS-IV. By setting the noise parameter to $p=1$, the expected Mermin value in Eq. \eqref{B4} reduces to the noiseless one under the strategy MS-III given in Ref. \cite{yx23pra}, which implies that the unbounded SSQN of the GHZ state can be realized by adopting the strategy MS‑IV assisted by local unitary operations in the noiseless scenario.

%%%%%%%%%%%%%%%%%%%%%%%%%%%%%%%%%%%%%%%%%%%%%%%%%%%%%%%%%%%%%%
\section{The proof of Theorem 3}

Starting from the initial W state in Eq. \eqref{W}, we can obtain the quantum state shared by Alice, Bob and Charlie$_k$ by applying a method analogous to that in Eqs. \eqref{A1}-\eqref{A2}. Then, the expected Mermin value for the phase-flip channel $\varepsilon_1$ under the strategy MS-V can be written as
\begin{eqnarray}\label{p M W}
	I_{\mathcal{M}}^{(k)}(\varepsilon_1)_{\mathrm{MS\text{-}V}}=\frac{P_{k}(\theta)}{2^{k-2}}\left(\cos^2\theta+\frac{2\sin^2\theta}{3}\right)+ \frac{\gamma_k(2p-1)^{k-1}}{2^{k-2}}\left(\frac{4\sin\theta \cos\theta }{3}\right),
\end{eqnarray}
where $P_{k}(\theta)=\prod\limits_{j=1}^{k-1}(1+\sqrt{1-\gamma_{j}^{2}})$ with $\theta$ being measurement parameter, $\gamma_k$ representing the measurement sharpness, and the noise parameter $p>1/2$.

Letting the expected Mermin value $I_{\mathcal{M}}^{(k)}(\varepsilon_1)_{\mathrm{MS\text{-}V}}>2$, we can define a sequence of sharpness parameters $\{\gamma_k(\theta)\}$ with the quantity $\epsilon>0$, which take the form
\begin{equation}\label{p gamma W}
	\gamma_{k}(\theta)=\left\{
	\begin{array}{ll}
		(1+\epsilon)(\frac{\tan(\theta)}{4}), & \hbox{$k=1$,} \\
		(1+\epsilon)\frac{2^{k-1}-P_{k}(\theta)(\cos^2\theta+\frac{2\sin^2\theta}{3})}{\frac{4\sin\theta \cos\theta}{3}(2p-1)^{k-1}} , & \hbox{$0\leq\gamma_{k-1}(\theta)\leq1$ ,} \\
		\infty, & \hbox{otherwise.}
	\end{array}
	\right.
\end{equation}
Since $0<\gamma_{k-1}(\theta)\leq1$, we can derive the ratio $\frac{\gamma_k(\theta)}{\gamma_{k-1}(\theta)}>2$. Moreover, we can obtain the sharpness $\gamma_{1}(\theta)=(1+\epsilon)(\frac{\tan(\theta)}{4})$ and the limit $\lim\limits_{\theta\rightarrow0^+}\gamma_{1}(\theta)=0$. Proceeding by induction, we assume the existence of $\theta_{k-1}$ such that on the interval $\theta\in(0,\theta_{k-1})$, all $\gamma_{i}(\theta)\in (0,1)$ and $\lim\limits_{\theta\rightarrow0^+}\gamma_{i}(\theta)=0$ for $i=1,2,\ldots,k-1$. For $P_k(\theta)$ defined on the interval $(0, \theta_{k-1})$, the limit of its derivative at $0^+$ is given by 
\begin{equation}\label{limit}
  \lim\limits_{\theta\rightarrow0^+}P^{'}_{k}(\theta)=\lim\limits_{\theta\rightarrow0^+}\sum_{j=1}^{k-1}\left(\frac{-\gamma_{j}(\theta)\gamma^{'}_{j}(\theta)}{\sqrt{1-\gamma^{2}_{j}(\theta)}}\frac{P_{k}(\theta)}{1+\sqrt{1-\gamma^{2}_{j}(\theta)}}\right)=0.
\end{equation}
Then, after applying L'H\^opital's rule and using the property $\lim\limits_{\theta\rightarrow0^+}P_{k}(\theta)=2^{k-1}$, we can obtain
\begin{align*}
	\lim\limits_{\theta\rightarrow0^+}\gamma_{k}(\theta)&=\lim\limits_{\theta\rightarrow0^+}
	(1+\epsilon)\frac{2^{k-1}-P_{k}(\theta)(\cos^2\theta+\frac{2\sin^2\theta}{3})}{\frac{4\sin\theta \cos\theta}{3}(2p-1)^{k-1}}\\
	&=\lim\limits_{\theta\rightarrow0^+}\frac{-P^{'}_{k}(\theta)\cos^2\theta+P_{k}(\theta)2\cos\theta \sin\theta-P^{'}_{k}(\theta)\frac{2\sin^2\theta}{3}-P_k(\theta)\frac{4\sin\theta \cos\theta}{3}}{\frac{4\cos2\theta}{3}(2p-1)^{k-1}}=0.
\end{align*}
Therefore, for any $n\in\mathbb{N}$, there exists a $\theta_n\in(0,1)$ such that $0<\gamma_{1}(\theta)<\gamma_{2}(\theta)<\cdots<\gamma_{n}(\theta)<1$ for all $\theta\in(0,\theta_n)$, which indicates that the unbounded sequential Mermin nonlocality sharing can be achieved in the presence of the phase-flip channel $\varepsilon_1$ by adopting the measurement strategy MS-V. 

In the case of the bit-flip channel $\varepsilon_2$ under our proposed strategy MS-VI assisted by local unitary operation $U_L$, the expected Mermin value for Alice, Bob, and Charlie$_k$ on the quantum state $|\psi\rangle_{W'}$ in Eq. \eqref{Wp} of the main text can be expressed as  
\begin{eqnarray}\label{p M Wx}
	I_{\mathcal{M}}^{(k)}(\varepsilon_2)_{\mathrm{MS\text{-}VI}}=\frac{P_{k}(\theta)}{2^{k-2}}(\cos^2\theta+\frac{2\sin^2\theta}{3})+ \frac{\gamma_k(2p-1)^{k-1}}{2^{k-2}}(\frac{4\sin\theta \cos\theta }{3}),
\end{eqnarray}
where $P_{k}(\theta)=\prod\limits_{j=1}^{k-1}(1+\sqrt{1-\gamma_{j}^{2}})$ with $\theta$ being measurement parameter, $\gamma_k$ denoting the measurement sharpness, and the noise parameter $p>1/2$. This result has the same expression as that for the phase-flip channel $\varepsilon_1$ under the strategy MS-V given in Eq. \eqref{p M W}. Therefore, under the bit‑flip channel, arbitrarily many independent Charlies can share Mermin nonlocality with a single Alice and a single Bob by adopting the strategy MS‑VI assisted by local operations on the standard W state.

For the bit-flip $\varepsilon_2$ channel under strategy MS-V, the phase-flip $\varepsilon_1$ channel under strategy MS-VI, and the depolarizing channel $\varepsilon_3$ under both the two strategies, the corresponding expected Mermin values can be written as
\begin{eqnarray}
	&&I_{\mathcal{M}}^{(k)}(\varepsilon_2)_{\mathrm{MS\text{-}V}}=I_{\mathcal{M}}^{(k)}(\varepsilon_1)_{\mathrm{MS\text{-}VI}}=\frac{(2p-1)^{k-1}}{2^{k-2}}P_{k}(\theta)\left(\cos^2\theta+\frac{2\sin^2\theta}{3}\right)+ \frac{\gamma_{k}}{2^{k-2}}\left(\frac{4\sin\theta \cos\theta }{3}\right),\label{C4}\\
    &&I_{\mathcal{M}}^{(k)}(\varepsilon_3)_{\mathrm{MS\text{-}V}}=I_{\mathcal{M}}^{(k)}(\varepsilon_3)_{\mathrm{MS\text{-}VI}}=\frac{p^{k-1}}{2^{k-2}}P_{k}(\theta)\left(\cos^2\theta+\frac{2\sin^2\theta}{3}\right)+ \frac{\gamma_{k} p^{k-1}}{2^{k-2}}\left(\frac{4\sin\theta \cos\theta }{3}\right),\label{C5}
\end{eqnarray}
where the noise parameter in Eq. \eqref{C4} is set to $p>1/2$. Similarly, we can define the corresponding sequences $\{\gamma_{k}(\theta)\}$. By induction, assume that there exists $\theta_{k-1}$ such that on the interval $(0, \theta_{k-1})$, $\gamma_i(\theta) \in (0,1)$ and $\lim\limits_{\theta\rightarrow0^+}\gamma_{k-1}(\theta)=0$ for $i = 1, 2, \ldots, k-1$. Under this hypothesis, $\gamma_{k}(\theta)$ does not vanish at the limit $\theta\rightarrow0^+$, which means that the tripartite Mermin nonlocality cannot be sequentially shared by arbitrarily many independent Charlies. 

Therefore, starting from an initial W state, for the phase-flip, bit-flip quantum channels, the unbounded sequential sharing of tripartite Mermin nonlocality can be realized by selectively adopting the strategies MS-V and MS-IV. But the unbounded SSQN breaks down under the influence of depolarizing noise under both the two measurement strategies. The proof of Theorem 3 is completed.

\hfill$\blacksquare$

Additionally, we analyze the SSQN in the noiseless scenario under our proposed measurement strategy MS-VI. By setting the noise parameter $p=1$,  the expected Mermin value in Eq. \eqref{p M Wx} reduces to the noiseless one under the strategy MS-V given in Ref.\cite{bz24pra}.This indicates that arbitrarily many independent Charlies can sequentially share the tripartite Mermin nonlocality with a single Alice and a single Bob in the noiseless case by adopting our proposed strategy MS‑VI assisted by local operations.

\section{The optimal values of measurement parameters for double-violation SSQN schemes }

In Sec. V of the main text, we analyze noise resilience of double-violation scheme for the CHSH inequality of the Bell state in the presence of channel noise, where where the measurement strategies MS-I and MS-II are employed and the expected CHSH values $I_{\mathcal{CHSH}}^{(1)}$ and $I_{\mathcal{CHSH}}^{(2)}$ are plotted as the functions of the measurement orientation $\theta$ as shown in Fig. \ref{f6}.

For the measurement strategy MS-I, the two expected CHSH values under the influence of phase-flip noise (with the noise parameter $p=0.9$) are the functions of measurement parameters $\theta$ and $\gamma_k$ according to Eq. \eqref{bellCHSH} in the main text, which can be expressed as
\begin{eqnarray}
	&&I_{\mathcal{CHSH}}^{(1)}(\varepsilon_1)_{\mathrm{MS\text{-}I}}=2\left[\gamma_{1}\sin(\theta)+\cos(\theta)\right],\label{D1}\\
	&&I_{\mathcal{CHSH}}^{(2)}(\varepsilon_1)_{\mathrm{MS\text{-}I}}=0.8\gamma_{2}\sin(\theta)+\cos(\theta)\left(1+\sqrt{1-\gamma_{1}^{2}}\right).\label{D2}
\end{eqnarray}
Using Eq. \eqref{17}, we can obtain the two sharpness parameters
\begin{eqnarray}
	&&\gamma_1=(1+\epsilon)\frac{1-\cos(\theta)}{\sin(\theta)},\label{D3}\\
	&&\gamma_2=(1+\epsilon)\frac{2-\cos(\theta)\left(1+\sqrt{1-\gamma_{1}^{2}}\right)}{0.8\sin(\theta)}.\label{D4}
\end{eqnarray} 
To achieve improved performance of the SSQN, we set the sharpness parameter for Bob$_2$ to $\gamma_2=1$. Then, for each given value of $\theta$, we find the optimal value of $\gamma_1$ as follows. Substituting Eq.~\eqref{D3} into Eq.~\eqref{D4}, we can obtain the optimal quantity $\epsilon$ by solving the equation $\gamma_2=1$. Subsequently, the optimal value of $\gamma_1$ can be derived by substituting the quantity $\epsilon$, which is obtained from Eq.~\eqref{D4}, into Eq.~\eqref{D3}. Therefore, for each value of $\theta$, we can derive the optimal values of parameters $\epsilon$ and $\gamma_1$ given that $\gamma_2=1$. The corresponding values for these parameters for Fig. \ref{f6}(a) are listed in Table ~\ref{bell_coefficients}, which can yield the expected CHSH values according to Eq. \eqref{D1} and \eqref{D2}. For other channels and strategies considered in Fig. \ref{f6}(b)-(d), we adopt the same parameter configuration.

\begin{table*}[b]
	\centering
	\caption{The optimal values of measurement parameters in the schemes for double-violation SSQN of the Bell state through noisy quantum channels with the noise strength $p=0.9$.}
	\label{bell_coefficients}
	\setlength{\tabcolsep}{2.5pt}
	\begin{tabular}{c*{14}{c}}
		\toprule
		$\theta$ & 0.01 & 0.05 & 0.1 & 0.15 & 0.2 & 0.25 & 0.3  & 0.35 & 0.4 & 0.45 & 0.5 & 0.55 & 0.6 & 0.65\\
		\hline
		$\epsilon$  & 7.3073 & 3.5089 & 2.3351 & 1.7397 & 1.3525 & 1.0711 & 0.8533  & 0.6775 & 0.5317 & 0.4080 & 0.3014 & 0.2084 & 0.1263 & 0.0533 \\
		\hline
		$\gamma_1$  & 0.0415 & 0.1127 & 0.1669 & 0.2059 & 0.2360 & 0.2602 & 0.2800 & 0.2966 & 0.3105 & 0.3223 & 0.3323 & 0.3410 & 0.3481 & 0.3549\\
		\hline\hline
	\end{tabular}
\end{table*}

We next investigate the optimal values of measurement parameters in the schemes for double-violation SSQN of the GHZ state evaluated by the Mermin inequality. In Fig. \ref{f7} of the main text, the expected Mermin values $I_{\mathcal{M}}^{(1)}$ and $I_{\mathcal{M}}^{(2)}$ are plotted as the functions of the parameter $\theta$ with the noise strength $p=0.8$ for different quantum channels.

For the measurement strategy MS-III, the two expected Mermin values under the influence of bit-flip noise are the functions of sharpness parameters $\theta$ and $\gamma_k$ according to Eq. \eqref{IM b G} of the main text, which can be written as
\begin{eqnarray}
&&	I_{\mathcal{M}}^{(1)}(\varepsilon_2)_{\mathrm{MS\text{-}III}}=2\theta(\gamma_{1}+1),\label{D5}\\
&&	I_{\mathcal{M}}^{(2)}(\varepsilon_2)_{\mathrm{MS\text{-}III}}=2\theta\left[0.3\gamma_{2}+\frac{1}{2}\left(1+\sqrt{1-\gamma_{1}^{2}}\right)\right].\label{D6}
\end{eqnarray}
Furthermore, based on Eq. \eqref{30}, we can derive  
\begin{eqnarray}
&&	\gamma_1=(1+\epsilon)\left(\frac{1}{\theta}-1\right),\label{D7}\\
&&	\gamma_2=(1+\epsilon)0.3^{-1}\left[\frac{1}{\theta}-\frac{1}{2}\left(1+\sqrt{1-\gamma_{1}^{2}}\right)\right].\label{D8}
\end{eqnarray}  
For each given value of $\theta$, the optimal $\epsilon$ can be obtained by substituting Eq.~\eqref{D7} into Eq.~\eqref{D8} and solving the equation by setting $\gamma_2 = 1$. Then, the optimal value of the sharpness parameter $\gamma_1$ can be subsequently obtained by substituting the quantify $\epsilon$ into Eq.~\eqref{D7}. 

The corresponding values of these parameters for Fig. \ref{f7}(a) are listed in Table~\ref{ghz_coefficients}, which can yield the expected Mermin values according to Eqs. \eqref{D5} and \eqref{D6}. For other noisy channel and strategies considered in Fig. \eqref{f7}(b)-(d), we adopt the same parameter configuration.

\begin{table*}[t]
	\centering
	\caption{The optimal values of measurement parameters in the schemes for double-violation SSQN of the GHZ state through noisy quantum channels with the noise strength $p=0.8$.}
	\label{ghz_coefficients}
	\setlength{\tabcolsep}{3.5pt}
	\begin{tabular}{c*{12}{c}}
		\toprule
		$\theta$ & 0.8 & 0.82 & 0.84 & 0.86 & 0.88 & 0.9  & 0.92 & 0.94 & 0.96 & 0.98 & 0.995 & 0.999\\
		\hline
		$\epsilon$  & 0.1123 & 0.2559 & 0.4324 & 0.6550 & 0.9447 & 1.3385 & 1.9071 & 2.8065 & 4.4712 & 8.8196 & 27.9752 & 92.6573 \\
		\hline
		$\gamma_1$  & 0.278067 & 0.2757 & 0.2728 & 0.2694 & 0.2652 & 0.2598 & 0.2528 & 0.2430 & 0.2280 & 0.2004 & 0.1456 & 0.0938\\
		\hline\hline
	\end{tabular}
\end{table*}

\twocolumngrid
\bibliographystyle{apsrevlong}
\bibliography{reference}

\end{document}